\documentclass[11pt,a4paper]{article}
\pdfoutput=1

\usepackage{jheppub}
\usepackage[T1]{fontenc}
\usepackage[english]{babel}
\usepackage{lmodern}
\usepackage{bbm}
\usepackage{bm}
\usepackage{color}
\usepackage{slashed}
\usepackage{graphicx}
\usepackage{braket}
\usepackage{bbold}
\usepackage{dsfont}

\DeclareGraphicsExtensions{.pdf,.png,.jpg}
\DeclareMathOperator{\arccot}{arccot}

\allowdisplaybreaks[1]
\def\simg{{\ \lower-1.2pt\vbox{\hbox{\rlap{$>$}\lower6pt\vbox{\hbox{$\sim$}}}}\ }}
\def\siml{{\ \lower-1.2pt\vbox{\hbox{\rlap{$<$}\lower6pt\vbox{\hbox{$\sim$}}}}\ }}

\makeatletter \@addtoreset{equation}{section} \makeatother

\newcommand*{\cf}{cf.\ }
\newcommand*{\ie}{i.e.\ }
\newcommand*{\eg}{e.g.\ }
\newcommand*{\Eq}{eq.~}
\newcommand*{\Eqs}{eqs.~}
\newcommand*{\myRef}{ref.~}
\newcommand*{\myRefs}{refs.~}

\title{Bound-state formation, dissociation and decays of 
	darkonium with potential non-relativistic Yukawa theory
	for scalar and pseudoscalar mediators}

\abstract{
	Dark matter models with light mediators featuring sizable interactions among dark particles
	enjoy an increasing attention in the model building community due to the elegance with which
	they can potentially explain the scaling relations governing galactic halos and clusters of galaxies. 
	In the present work we continue our study of such models using non-relativistic and potential
	non-relativistic effective field theories (NREFTs and pNREFTs) and explore the properties of a 
	Yukawa-type model with scalar and pseudoscalar interactions between a 
	low-energetic scalar mediator and heavy dark matter fermions.
	In particular, we make first steps towards the formulation of such theories at finite temperature
	by providing the thermal bound-state formation rate and the thermal break-up of bound states from the self-energies of the dark-pair fields, that interact with the thermal environment. We estimate numerically bound-state effects on the dark matter energy density, that provide up to a $35\%$ correction depending on the relative size of the model couplings.
}

\author[a]{Simone Biondini}
\author[b]{and Vladyslav Shtabovenko}

\emailAdd{simone.biondini@unibas.ch}
\emailAdd{v.shtabovenko@kit.edu}

\affiliation[a]{Department of Physics, University of Basel, Klingelbergstr. 82, CH-4056 Basel, Switzerland} 
\affiliation[b]{Institut für Theoretische Teilchenphysik (TTP), Karlsruhe Institute of Technology (KIT), Wolfgang-Gaede-Straße 1, 76131 Karlsruhe, Germany}

\preprint{TTP21-058, P3H-21-100}
\arxivnumber{2112.10145}

\begin{document}
\maketitle

\section{Introduction and motivation}
The quest for uncovering the true nature of dark matter (DM) comes as a compelling and challenging endeavor for both particle physics and cosmology communities. On the one hand, gravitational effects provide a strong evidence of the existence of DM, while possible alternative explanations seem inconclusive. On the other hand, we still lack the much anticipated discovery of signals of dark particles at colliders or experimental facilities dedicated to direct as well as indirect DM searches.

The non-observation of any DM signal has resulted in cornering many minimal DM models.
In particular, models comprising weakly interacting massive particles (WIMPs) are under strong tension with current experimental observations~\cite{Arcadi:2017kky}. This naturally prompts us to consider the existence of richer dark sectors, containing multiple dark particles, \cf \eg \myRefs\cite{Bertone:2004pz,DeSimone:2016fbz,Petraki:2013wwa}.
 The dark particles can experience their own hidden forces, which are partially or entirely secluded from the Standard Model (SM) sector. Furthermore, the existence of light force  mediators with masses much smaller than that of the actual DM particles may affect the DM dynamics in many ways. A straightforward consequence is that DM may experience sizable self-interactions that depend on the relative velocity among dark particles and can provide a dynamical explanation for the scaling relations governing galactic halos and clusters of galaxies~\cite{Foot:2013nea,Foot:2013lxa,Foot:2013uxa,Markevitch:2003at,Randall:2007ph,Kahlhoefer:2013dca,Harvey:2015hha,Kaplinghat:2015aga,Laha:2015yoa}. In this sense, from a phenomenological point of view a next-to-minimal dark sector is more than welcome. 
 
 Another remarkable property of DM models with light mediators is the possibility to have bound states within the dark sector~\cite{Feng:2009mn,Detmold:2014qqa,vonHarling:2014kha}. Depending on the details of the model, both stable and unstable bounds state may form. Under the conditions, that the dark particles have reached a thermal equilibrium in the early universe and the DM relic density is generated through the well-known freeze-out mechanism, bound-state formation and dissociation can become important for an accurate determination of the present-day energy density. Indeed, whenever bound states are formed, and not effectively dissociated or melted away in the thermal plasma, they provide an additional process for the depletion of DM particles in the early universe.\footnote{Bound-state effects on non-thermally produced DM particles have been considered in \myRefs\cite{Biondini:2020ric,Bollig:2021psb,Garny:2021qsr}.}
 
 A quantitative estimation of bound-state effects on the DM relic density is model dependent. Most of the past and recent literature is focused on vector mediators, that resulted in complementary and diversified approaches to calculate formation cross sections at zero~\cite{Feng:2009mn,vonHarling:2014kha,Beneke:2016ync,Liew:2016hqo,Cirelli:2007xd,Mitridate:2017izz,Harz:2019rro}  and at finite temperature~\cite{Kim:2016kxt,Biondini:2017ufr,Biondini:2018pwp,Biondini:2018ovz,Binder:2018znk,Biondini:2019int, Binder:2019erp,Biondini:2019zdo,Binder:2020efn}. A more careful look at scalar mediators has been initiated only recently~\cite{Wise:2014jva,Petraki:2015hla,An:2016kie,Biondini:2018xor,Oncala:2018bvl,Harz:2019rro,Oncala:2019yvj,Oncala:2021tkz,Oncala:2021swy}. Even without the addition of scalars in the dark sector, it has been shown that if the dark particles couple to the SM Higgs boson, and DM is sufficiently heavier than the SM scalar, Sommerfeld enhancement and bound-state formation are relevant~\cite{Harz:2017dlj,Beneke:2014gja}. The in-vacuum bound-state formation  has been scrutinized for both neutral and charged scalar mediators and, in the latter case, it was found that fast monopole interactions can dramatically change the relic density estimation even for small values of the couplings~\cite{Oncala:2019yvj,Oncala:2021tkz,Oncala:2021swy}. 
 
 As for the treatment of the bound-state dynamics, one of the main approaches adopted so far relies on the Bethe-Salpeter~\cite{Salpeter:1951sz} equation for the wave functions, and the reduction thereof to a Schrödinger equation using the instantaneous
approximation in the non-relativistic regime~\cite{Petraki:2015hla}. The building blocks of this framework are relativistic four-point Green functions of single DM particles at zero temperature, whereas~\cite{Binder:2018znk} features a finite-temperature version of this formalism. A complementary approach, that we adopt in this work, is to exploit the technology of non-relativistic effective field theories (NREFTs). Our main goal is to provide quantum field theoretic tools for computing thermal processes involving bound states, or in general heavy pairs interacting through a potential. It is easy to see that the problem at hand comes as a multi-scale system. One finds three typical and presumably well-separated scales relevant for the non-relativistic dynamics, which are $M \gg Mv \gg Mv^2$, where $M$ is the DM particle mass, while $v$ stands for its typical velocity in a bound state. A Coulombic bound state satisfies $v \sim \alpha$, with $\alpha$ being the relevant coupling constant. Furthermore, our system comprises the mediator mass $m$, which is assumed to be much smaller than $M$, and the thermal scales, most notably the temperature of the early universe. Whenever the scalar mediator can interact with other light degrees of freedom in the dark sector, or with itself via a self-coupling, a thermal mass $m_T$ can be generated. In this case it must be added to the hierarchy of scales mentioned beforehand.
 
 Bound states in general exhibit a strong sensitivity to thermal scales, and it is rather useful to make an analogy between a well-studied system that arises in QCD, namely heavy quarkonium in a quark-gluon plasma, and heavy DM pairs in the early universe. In the following we choose to call a bound state made of two DM particles a \emph{darkonium}. Two processes are at play for the dissociation of quark-antiquark bound states in the medium, which are the gluo-dissociation~\cite{Kharzeev:1994pz} and the dissociation by inelastic parton scattering~\cite{Grandchamp:2001pf}. The former describes a situation, where a thermal gluon hits the quark-antiquark pair in a singlet configuration and, if sufficient energy is available,  breaks it into an unbound color-octet state. In this work we will address the analogue process induced by a thermal scalar being responsible for the breaking of a darkonium into a scattering state. The second dissociation process comes as a $2 \to 2$ scattering reaction, where the thermal particles in the bath exchange energy with the heavy quarkonium through a gluon exchange, turning a bound-state into an unbound one. A distinctive feature of an EFT approach is the availability of rigorous power counting rules that can be used to show that the gluo-dissociation process is dominant over the inelastic parton scattering for $E >m_D$~\cite{Laine:2006ns,Brambilla:2008cx,Escobedo:2008sy}, where $m_D \sim g_s T$ is a thermal mass for the gluon called Debye mass and $E \sim Mv^2$ defines the ultra-soft scale.

 It is now clear that in order to follow the intricate dynamics of bound states in a thermal medium, an EFT approach constitutes a powerful tool that  (i) allows scrutinizing the various arrangements of the energy scales and (ii) can be employed to devise the proper field theory to carry out the derivation of cross sections and decay widths by using the relevant degrees of freedom at the scale of interest. We shall attack this problem by using a recently proposed EFT for non-relativistic fermion-antifermion pairs interacting via a scalar mediator. Building upon the well-known NREFTs and potential NREFTs (pNREFTs) of QED and QCD~\cite{Caswell:1985ui,Bodwin:1994jh,Pineda:1997bj,Brambilla:1999xf,Brambilla:2004jw} as well as scalar Yukawa theory
 \cite{Luke:1996hj,Luke:1997ys}, we developed
 potential non-relativistic Yukawa theory (pNRY)~\cite{Biondini:2021ccr} that is well suited to address the questions at hand. The relevant degrees of freedom of pNRY are heavy fermion-antifermion pairs and ultra-soft scalars. In the first paper on the subject, we have applied pNRY to compute the darkonium spectrum and the bound-state formation cross section at zero temperature. Here we make a first step towards the generalization of pNRY at finite temperature. We shall use the so-obtained EFT to derive the thermal cross section for bound-state formation and the reverse process, namely the dissociation rate via the absorption of a thermal scalar from the medium. The latter process is a genuine finite temperature effect, and we will show how it can be recast as a convolution of the thermal distribution of the scalar particles and an in-vacuum dissociation cross section, without the need of Boltzmann-like prescriptions. 
 Our pNREFT can be also used to treat pair annihilations, which are known to be key ingredients for DM freeze-out calculations. Here, the Sommerfeld-enhanced cross section and bound-state decay width will emerge naturally. Our program closely follows the successful strategy outlined in the case of heavy quarkonium at finite temperature~\cite{Brambilla:2008cx,Brambilla:2010vq,Brambilla:2011sg}. It is worth mentioning that an EFT approach to search for DM in atomic-spectroscopy measurements has been recently developed for different types of force mediators, including scalar and pseudoscalar ones,  in ref.\cite{Frugiuele:2021bic}. 

In this work we focus on a particular hierarchy of scales, that  reads
\begin{equation}
    M \gg M \alpha \sim M v \gg \pi T \sim M \alpha^2 \sim M v^2 \gg m
    \label{hiearchy}
\end{equation}
and enables us to use pNRY as a starting point, as we explain in section~\ref{sec:pNRY_T}. This hierarchy of scales is motivated by two main aspects. First, temperatures of order of or smaller than the binding energy are interesting for DM phenomenology, since a depletion of DM pairs in the form of bound states is more efficient in such a temperature window. Indeed, due to a suppression of the dissociation process, bound-state formation becomes more relevant and opens up an additional channel for the pair annihilation. Second, as guided by the potential non-relativistic QCD (pNRQCD) at finite temperature and the inherent power counting, it is expected that the bound-state formation (dissociation) via scalar emission (absorption) dominates over $2 \to 2$ scattering processes with plasma constituents. This latter effect will be addressed in a future study on the subject, as it requires the derivation of thermal self-energies for the scalar field at NLO.\footnote{A finite temperature treatment of NLO effects and the connection to the interactions among heavy pairs and bound states for a DM model with a vector mediator has been carried out in \myRef\cite{Binder:2020efn,Binder:2021otw}.} A similar investigation for the same  hierarchy of scales \eqref{hiearchy}, however in the case of an Abelian DM model with a vector mediator, is in preparation \cite{B_and_NAG_1}.  

The structure of the paper is as follows. In section~\ref{sec:pNRY_T}
 we introduce pNRY$_{\gamma_5}$, a variety of pNRY featuring scalar and pseudoscalar Yukawa interactions, and discuss the thermal propagators of the heavy pair and the scalar mediator as the main ingredients for the finite temperature treatment. Then, in section~\ref{sec:pNRY_application}, the bound-state formation and dissociation is computed in pNRY$_{\gamma_5}$ starting from the self-energies of the heavy-pair field. The thermal rates are naturally obtained in terms of quantum mechanical matrix elements, that are analytically simplified in the Coulombic regime for the first $nS$ states. Annihilations of heavy pairs in pNRY$_{\gamma_5}$ are discussed in section~\ref{sec:pNRYann}. Next, section~\ref{sec:pheno} is devoted to a phenomenological study of the DM energy density, where the bound-state formation, bound-state decay width and the thermal width from scalar dissociation enter as key ingredients. Conclusions are offered in section~\ref{sec:concl}, while some technical details underlying our results are collected in the appendices.

 \section{Non-relativistic EFTs for a scalar mediator}
\label{sec:pNRY_T}
In this section we introduce the DM model featuring a light scalar force mediator between heavier DM particles, in particular fermions. Such model Lagrangian will be our fundamental theory, from which we can construct towers of low-energy theories by integrating out energy scales. Next, we proceed with the discussion of the pNRY Lagrangian, and we introduce the thermal propagators of the heavy-pair field and the scalar particle. 
\subsection{Dark matter model}
 We assume DM to be a Dirac fermion singlet under the SM gauge group that couples to a scalar particle via Yukawa-type interactions. The Lagrangian density of the model reads~\cite{Pospelov:2007mp,Kaplinghat:2013yxa} 
\begin{equation}
    \mathcal{L}= \bar{X} (i \slashed{\partial} -M)X + \frac{1}{2} \partial_\mu \phi \, \partial^\mu \phi -\frac{1}{2}m^2 \phi^2 - \frac{\lambda}{4!} \phi^4 -  \bar{X} (g + ig_5 \gamma_5)   X \phi +\mathcal{L}_{\hbox{\scriptsize portal}} \, ,
    \label{lag_mod_0}
    \end{equation}
where $X$ is the DM Dirac field and $\phi$ is a real scalar field. The scalar self-coupling is denoted with $\lambda$, whereas the scalar and pseudo-scalar couplings are $g$ and $g_5$ respectively. The mass of the scalar mediator $m$ is assumed to be much smaller than the DM particle mass $M$, $m \ll M$. In our work, we adopt a simplified model realization, where the issues of the fermion mass generation and of the gauge group governing the dark sector are not addressed (\cf \eg\cite{Kahlhoefer:2015bea,Duerr:2016tmh} for a simplified model with two mediators, scalar and vector, where the gauge invariance and spontaneous symmetry breaking in the dark sector is fully accounted for). Our aim is to consider the Lagrangian given in \Eq\eqref{lag_mod_0} as one of the simplest representatives for minimal DM models~\cite{Kaplinghat:2013yxa,DeSimone:2016fbz} with a light scalar mediating interactions between DM particles. It is worth mentioning that the model can be much more involved, and can be extended to have a richer set of interaction terms~\cite{Wise:2014jva,Kahlhoefer:2017umn,Oncala:2018bvl}. For example, an interaction of the form $\rho_\lambda \phi^3$ can be foreseen, that is responsible for an additional bound state formation process~\cite{Oncala:2018bvl,Oncala:2019yvj}. 

Then, $\mathcal{L}_{\hbox{\scriptsize portal}}$ comprises the interactions between the scalar $\phi$ and other degrees of freedom that can belong to the SM sector or to the dark sector. One of the most common realizations of such a portal involves interactions with the SM Higgs boson. Portal interactions are welcome in order to introduce a mechanism that allows $\phi$ particles to decay and deplete their population. Indeed, the light scalar particles $\phi$ are abundant in the early universe and a substantial population survives after the freeze-out of the dark fermion~\cite{Kaplinghat:2013yxa,DelNobile:2015uua}.\footnote{As for the Higgs portal, the interaction terms with the scalar mediator $\phi$  read, before electroweak symmetry breaking, as  $ \mathcal{L}_{\textrm{portal}} = - a \phi \mathcal{H}^\dagger \mathcal{H} - b \phi^2 \mathcal{H}^\dagger \mathcal{H}$; $\mathcal{H}$ is the SM Higgs doublet. The smallest portal couplings that ensure a thermalization between the SM and dark sector, and allow for the decay of the scale $\phi$ before BBN, are typically much smaller than the electroweak gauge couplings, see \eg \cite{Kaplinghat:2013yxa,Enqvist:2014zqa}. We have explicitly verified this for the assumed range of numerical values for our model parameters $g$, $g_5$ and $M$. This means that the portal interactions can be safely neglected in our EFT construction and the corresponding matching calculations.}
A richer portal sector also allows alleviating and often removing the tension with the experiments for the model, if one considers only the interactions of the scalar $\phi$ with the Higgs boson~\cite{Kaplinghat:2013yxa,Kainulainen:2015sva}.

However, when dealing with a thermal environment, many interactions between the scalar mediator and other plasma constituents may endanger the assumed hierarchy of scales. In particular, a thermal mass of the form $m_T \sim g' T$ is generated, that can become larger than the in-vacuum mass or the typical ultrasoft scale $M \alpha^2$. Moreover, in the case of the bound-state formation process, when the scalar mass exceed the difference between an above-threshold scattering state and a bound-state, \cf \Eq\eqref{diff_scat_bound}, the $1 \to 2$ radiative formation process becomes kinematically forbidden. For example, if one only considers the scalar self interaction, a thermal scalar mass  $m_T = \sqrt{\lambda/12}T$ is generated, that translates into the condition $T \ll M \alpha^2  \sqrt{12/\lambda} $ to attain the hierarchy of scales given in \Eq\eqref{hiearchy}. We defer the scrutiny of the impact of thermal masses, and a thermal width, of the scalar propagator to future works on the subject. Since we included a fermion-pseudoscalar interaction, we denote the following towers of EFTs with NRY$_{\gamma_5}$ and pNRY$_{\gamma_5}$ in order to make clear that the corresponding low-energy theories differ from the ones obtained in the case of the sole scalar interaction~\cite{Biondini:2021ccr}.

To ensure the correctness of the results presented in the next sections we make use of computer tools
that were already employed in our previous work on the subject~\cite{Biondini:2021ccr}. Calculations within the fully relativistic DM model given in \Eq\eqref{lag_mod_0} can be automatized using  \textsc{FeynArts}~\cite{Hahn:2000kx}, \textsc{FeynRules}~\cite{Alloul:2013bka} and \textsc{FeynCalc}~\cite{Mertig:1990an,Shtabovenko:2016sxi,Shtabovenko:2020gxv}. The matching between the full theory and the corresponding non-relativistic EFTs is done using \textsc{FeynOnium}~\cite{Brambilla:2020fla}, while we employ \textsc{QGRAF}~\cite{Nogueira:1991ex} interfaced to \textsc{FeynCalc} via \textsc{FeynHelpers}~\cite{Shtabovenko:2016whf} to generate the corresponding Feynman diagrams. Together with extensive pen and paper cross checks, this approach greatly facilitates the task of avoiding typos and unintentional mistakes.

\subsection{\texorpdfstring{NRY$_{\gamma_5}$}{NRY5}}
\label{sec:nry}
We now want to proceed to the construction of the low-energy theory relying on the hierarchy of scales in \Eq\eqref{hiearchy}. 
As highlighted and extensively discussed in the existing studies of heavy quarkonia as well as hydrogen and muonic atoms at finite temperature~\cite{Brambilla:2008cx,Escobedo:2008sy,Escobedo:2010tu}, one can first integrate out the in-vacuum scales $M$ and $M \alpha$. Subsequently, all the smaller scales, including the thermal ones, can be set to zero. Let us stress that in practice the matching and the derivation of the low-energy theories is insensitive to the thermal scales. Therefore, we can first integrate out the scale $M$ to obtain NRY$_{\gamma_5}$ and then do the same for the scale $M \alpha$ thus arriving at pNRY$_{\gamma_5}$. We refer to \myRef\cite{Biondini:2021ccr} for further details on the construction of these two EFTs.

NRY$_{\gamma_5}$ is well suited to describe non-relativistic fermions and antifermions interacting with a soft scalar. In particular, fermion-antifermion annihilations into light mediators are accounted for by four-fermion operators. Schematically, the NRY$_{\gamma_5}$ Lagrangian reads
\begin{eqnarray}
\mathcal{L}_{\hbox{\tiny NRY$_{\gamma_5}$}} = \mathcal{L}^{\textrm{bilinear}}_{\psi} + \mathcal{L}^{\textrm{bilinear}}_{\chi} + \mathcal{L}_{\textrm{4-fermions}}  + \mathcal{L}_{\textrm{scalar}} \, ,
\label{NRY_scheme}
 \end{eqnarray}
 where $\psi$ ($\chi$) is the Pauli field that annihilates (creates) a heavy fermion, while all the scalar particles have energy and momenta much smaller than $M$. The inclusion of the pseudoscalar interaction as in \Eq\eqref{lag_mod_0} introduces new operators in the bilinear sector with respect to NRY, whereas the set of four-fermion operators and $\mathcal{L}_{\textrm{scalar}}$ remain the same. As for the modification of the bilinear sector, we find that the leading fermion-pseudoscalar interaction is suppressed with respect to the leading fermion-scalar interaction by $k/M$, where $k$ is the soft or ultra-soft momentum carried by the field $\phi$. The fermion bilinear at order $1/M$ with the matching coefficients set to their tree-level values reads
\begin{equation}
     \mathcal{L}^{\textrm{bilinear}}_{\psi} = \phantom{+} \psi^\dagger \left( i \partial_0  - \,  g\phi +  g_5 \frac{\sigma \cdot [\bm{\nabla} \phi]}{2M} - g_5^2 \frac{\phi^2}{2M} + \frac{\bm{\nabla}^2}{2 M} \right) \psi \, .
     \label{bilinear_psi_5}
\end{equation}
As for the antifermion fields one finds
\begin{equation}
     \mathcal{L}^{\textrm{bilinear}}_{\chi} = \phantom{+} \chi^\dagger \left( i \partial_0  + \,  g\phi +  g_5 \frac{\sigma \cdot [\bm{\nabla} \phi]}{2M} + g_5^2 \frac{\phi^2}{2M}  - \frac{\bm{\nabla}^2}{2 M} \right) \chi \, .
     \label{bilinear_chi_5}
\end{equation}
Some details on the derivation of the above fermion-bilinear Lagrangians are given in appendix~\ref{sec:appendix-nry}. The operators proportional to one power of $g_5$ are parity violating, and the notation $[\bm{\nabla} \phi]$ stands for the derivative acting on the scalar field only. The operators proportional to $g_5^2$ and involving two powers of the scalar fields were absent in NRY. Our result for the bilinear sector with a pseudoscalar interaction agrees with the findings in \myRef\cite{Platzman:1960dqa}.

Most notably, the presence of the pseudoscalar coupling generates non-vanishing matching coefficients for the leading dimension-6 four-fermion operators. In this paper we are mostly interested in such leading  order  modification  to  the  four-fermion  sector,  whereas  the  correction  to  the  bilinear sector, and the corresponding ones in the pNRY$_{\gamma_5}$, will not be carried out in full details. The two independent dimension-6 operators read~\cite{Bodwin:1994jh}  
\begin{eqnarray}
( \mathcal{L}_{\textrm{4-fermions}} )_{d=6} = \frac{f(^1S_0)}{M^2} \psi^\dagger \chi \, \chi^\dagger \psi + \frac{f(^3S_1)}{M^2} \psi^\dagger \, \bm{\sigma} \, \chi \cdot \chi^\dagger \, \bm{\sigma} \, \psi \, ,
\label{dimension_6_lag}
\end{eqnarray}
and the corresponding matching coefficients are found to be 
\begin{equation}
     {\rm{Im}}[f(^1S_0)] = 2  \pi \alpha \alpha_5 \, , \quad  {\rm{Im}}[f(^3S_1)] = 0 \, .
      \label{match_sca_dim_6}
\end{equation}
The spectroscopy notation is borrowed from NRQED/NRQCD, so that one can classify  the annihilations in terms of the total spin $S$ of the pair, the relative angular momentum $L$ and the total angular momentum $J$, by writing  $^{2S+1}L_J$. The pseudoscalar interaction modifies the matching coefficients of the velocity suppressed operators as well. These dimension-8 operators are 
\begin{eqnarray}
( \mathcal{L}_{\textrm{4-fermions}} )_{d=8} &=& \frac{f(^1P_1)}{M^4} \mathcal{O}(^1P_1) + \frac{f(^3 P_0)}{M^4} \mathcal{O}(^3 P_0) + \frac{f(^3 P_1)}{M^4} \mathcal{O}(^3 P_1)  \nonumber 
\\
 &+& \frac{f(^3 P_2)}{M^4} \mathcal{O}(^3 P_2) + \frac{g(^1 S_0)}{M^4} \mathcal{P}(^1 S_0) + \frac{g(^3 S_1)}{M^4} \mathcal{P}(^3 S_1)
 \nonumber 
 \\
  &+& \frac{g(^3 S_1,^3 D_1 )}{M^4} \mathcal{P}(^3 S_1, ^3 D_1) + \cdots \, , 
\label{dimension_8_lag}
\end{eqnarray}
where we refer to \myRef\cite{Bodwin:1994jh} for their explicit definitions. The matching coefficients, that generalize our result for the sole scalar interaction in \myRef\cite{Biondini:2021ccr}, read
\begin{subequations}
\begin{eqnarray}
   && {\rm{Im}}[f(^1P_1)] =  {\rm{Im}}[f(^3P_1)] = 0  \, ,
    \label{match_sca_dim_8_a}
   \\
   &&   {\rm{Im}}[f(^3P_0)] = \frac{\pi}{6} (5 \alpha -  \alpha_5)^2 \, , \quad  {\rm{Im}}[f(^3P_2)] = \frac{\pi}{15}  (\alpha + \alpha_5)^2 \, ,
    \label{match_sca_dim_8_b}
   \\
   && {\rm{Im}}[g(^1S_0)] =  -\frac{8 \pi}{3} \alpha \alpha_5 \, , \quad  {\rm{Im}}[g(^3S_1)]  = {\rm{Im}}[g(^3S_1, ^3D_1)] = 0 \, .
    \label{match_sca_dim_8}
\end{eqnarray}
\end{subequations}
 It is worth to mention that the vanishing of $f(^3 S_1)$ and also $f(^1 P_1)$ can be understood using symmetry arguments.
Although the pseudoscalar Yukawa interaction violates parity, it still preserves the charge conjugation symmetry. The conservation of the charge conjugation selects particular combinations of spin and angular momentum of the annihilating fermion-antifermion pair into two scalars.\footnote{The $C$ quantum number for a bound state of two fermions is $(-1)^{L+S}$, whereas a scalar field is a parity eigenstate, we have $C=1^n$ for $n$ scalar particles in the final state. This would select only $C=1$ darkonium states, which are ${}^1 S_0$ and ${}^3 P_J$. Additional light fermion species, that couple to the scalar field, can lift this condition and allow nonvanishing matching coefficients.  }

We conclude this section by calculating the non-relativistic annihilation cross section for the process $X \bar{X} \to \phi \phi$ without any kind of resummation of the soft physics, which implies that we are solely sensitive to the hard energy modes at the scale  $M$. In order to compare to the results in the literature, we average over spin polarizations of the incoming fermion and antifermion.
The cross section then reads
\begin{eqnarray}
    \sigma_{ \textrm{ann}} =  \frac{2{\rm{Im}}[\mathcal{M}(\psi\chi \to \psi\chi)]}{(2 s_\psi+1)(2 s_\chi+1)  v_{\textrm{rel}} } \, ,
    \label{X_section_NR}
    \end{eqnarray}
	where we choose the heavy fermion fields to be normalized non-relativistically as is customary in NRQCD~\cite{Bodwin:1994jh}. The relative velocity in the center-of-mass frame is given by $v_{\textrm{rel}}=|\bm{v}_\psi-\bm{v}_\chi|=2 v$. The imaginary part of the non-relativistic amplitude in the numerator of \Eq\eqref{X_section_NR} can be easily computed using the four-fermion Lagrangians given in \Eqs\eqref{dimension_6_lag} and \eqref{dimension_8_lag}, and the matching coefficients in \Eqs\eqref{match_sca_dim_6} and \eqref{match_sca_dim_8_a}-\eqref{match_sca_dim_8}. We obtain 
    \begin{eqnarray}
  \sigma_{\textrm{ann}} v_{\textrm{rel}} &=& \frac{1}{M^2} \left\lbrace  \left( {\frac{1}{3}{\rm{Im}}[f(^3P_0)]} + {\frac{5}{3}{\rm{Im}}[f(^3P_2)]}  + {\rm{Im}}[g(^1S_0)]  \right) \frac{v_{\textrm{rel}}^2}{4} + {\rm{Im}}[f(^1S_0)] \right\rbrace
  \nonumber\\
 & =& \frac{ 2 \pi \alpha \alpha_5}{M^2} + \frac{\pi}{24 M^2} (9 \alpha^2 + \alpha_5^2  -18 \alpha \alpha_5) v_{\textrm{rel}}^2  \, .
    \label{NR_hard_cross_section_S}
\end{eqnarray}
 The result agrees with the literature~\cite{Wise:2014jva,An:2016kie} as for the terms proportional to $\alpha^2 v_{\textrm{rel}}^2$ and $\alpha_5^2 v_{\textrm{rel}}^2$ and the velocity independent term, whereas the term proportional to the product of the couplings at order $v_{\textrm{rel}}^2$ is new. To the best of our knowledge, the $\alpha \alpha_5 v_{\textrm{rel}}^2$-piece was not included in former derivations. An analogous recasting of DM annihilations, as in the first line of \Eq\eqref{NR_hard_cross_section_S}, in the context of supersymmetric models can be found in \myRef\cite{Beneke:2012tg}. 

\subsection{\texorpdfstring{pNRY$_{\gamma_5}$}{pNRY5}}
\label{sec:pNR}

Having obtained NRY$_{\gamma_5}$, the next natural step is to derive the corresponding EFT at the ultra-soft scale in order to calculate processes relevant to this work, namely the bound-state formation cross section, the bound-state dissociation width and the bound-state decays. To this aim, we work with the pNRY Lagrangian~\cite{Biondini:2021ccr} where the degrees of freedom are heavy fermion-antifermion pairs and low-energetic scalars. In the following we briefly summarize its form and elucidate on the modifications that occur when the pseudo-scalar interaction is considered. 

In order to obtain NRY$_{\gamma_5}$, we integrated out the hard scale $M$.
The next scale one has to integrate out according to the scale hierarchy in \Eq\eqref{hiearchy} is $M \alpha$, or the inverse of the typical distance between $X$ and $\bar{X}$ in a pair.\footnote{It is worth pointing out that the physical degrees of freedom that are removed when going to pNRY, or pNRY$_{\gamma_5}$, are soft and potential scalars with energies and momenta larger than $M \alpha^2$. Hence, in the zero temperature case only ultrasoft scalars having $E \sim |\bm{p}| \sim M \alpha^2$ remain.
As far as non-relativistic fermions are concerned, we are left with pairs having $E \sim M \alpha^2$ and $|\bm{p}| \sim M \alpha$.} As long as we assume that the mass of the scalar satisfies $m \ll M\alpha$, we find ourselves in a Coulomb-like regime, where the velocity of the particles in a bound state scales as $v \sim \alpha$. As explained in section~\ref{sec:nry}, we can set the thermal scales to zero in the matching between NRY$_{\gamma_5}$ and pNRY$_{\gamma_5}$. In practice, one projects the NRY$_{\gamma_5}$ Hamiltonian onto the particle-antiparticle sector via 
\begin{equation}
\int d^3 \bm{x}_1 d^3 \bm{x}_2 \, \varphi_{ij}(t,\bm{x}_1, \bm{x}_2) \psi^\dagger_i (t,\bm{x}_1) \chi_j (t,\bm{x}_2) | \phi_{\hbox{\tiny US}}\rangle \, ,
\label{proj_fock}
\end{equation}  
where $i,j$ are Pauli spinor indices, while the state $| \phi_{\hbox{\tiny US}}\rangle$ contains no heavy particles or antiparticles but an arbitrary number of scalars with energies much smaller than $M\alpha$, including thermal scales. Here $\varphi_{ij}(t,\bm{x}_1, \bm{x}_2)$ is a bilocal wave function field representing the $X\bar{X}$ system. 
The \emph{multipole-expanded} Lagrangian written in terms of the relative distance of the pair $\bm{r}=\bm{x}_1-\bm{x}_2$ and its center-of-mass coordinate $\bm{R}=\bm{x}_1+\bm{x}_2$  reads
\cite{Biondini:2021ccr}
\begin{align}
     L_{\hbox{\tiny pNRY$_{\gamma_5}$}}  &= \int d^3 \bm{r} \, d^3 \bm{R}  \, \varphi^\dagger(\bm{r},\bm{R},t) \left\lbrace  i \partial_0 +\frac{\bm{\nabla}^2_{\bm{r}}}{M} +\frac{\bm{\nabla}^2_{\bm{R}}}{4M} + \frac{\bm{\nabla}^4_{\bm{r}}}{4 M^3} - V(\bm{p},\bm{r},\bm{\sigma}_1,\bm{\sigma}_2)\right. 
     \nonumber \\
     & \left. - 2 g \phi(\bm{R},t) -g\frac{ r^i r^j }{4}  \left[ \nabla_R^i \nabla_R^j \, \phi (\bm{R},t)   \right] -   g \phi(\bm{R},t) \frac{\bm{\nabla}^2_{\bm{r}}}{M^2}    \right\rbrace \varphi(\bm{r},\bm{R},t)
     \nonumber
     \\
     &+ \int d^3 \bm{R} \left[ \frac{d_1}{2} \partial^\mu \phi \, \partial_\mu \phi - d_2 \frac{m^2}{2} \phi^2 + \frac{ d_3}{4!} \phi^4 + \frac{d_4}{M^2}  (\partial^\mu \phi) \partial^2 (\partial_\mu \phi) + \frac{d_5}{M^2}  (\phi \partial^\mu \phi) (\phi \partial_\mu \phi)  \right],
     \nonumber
     \\
     \label{pNREFT_sca_0}
 \end{align}
 where the square brackets in the second line of \eqref{pNREFT_sca_0} indicate that the spatial derivatives act on the scalar field only, which has to be understood as multipole expanded in the last line of \Eq\eqref{pNREFT_sca_0} as well.
 The typical size of the two coordinates is given by $\bm{r} \sim 1/(M \alpha)$ and $\bm{R} \sim 1/(M \alpha^2)$.
 The multipole expansion for the scalar fields in $\bm{r} \ll \bm{R}$ ensures that they carry only ultra-soft energies and momenta. To avoid cluttering the notation we suppress the spin indices of the bilocal fields that are contracted with each other. Each term in the pNRY$_{\gamma_5}$ Lagrangian has a well-defined scaling. The time derivative scales as $\partial_0 \sim M \alpha^2 \sim T$, the inverse relative distance and the corresponding derivative obey ${\bm{r}}^{-1}, \bm{\nabla}_{\bm{r}} \sim M \alpha$, whereas the scalar field and the center-of-mass derivative satisfy $g\phi, \bm{\nabla}_{\bm{R}} \sim M \alpha^2 \sim T$. Indeed, the dynamical scales active in the so-obtained EFT are the ultrasoft scale \textit{and} the temperature. 
 
 The potential is understood as a matching coefficient and is organized as an expansion in $\alpha(M)$ and $\lambda(M)$, as well as $1/M$, the  coupling $\alpha(1/r)$ and the relative distance $r$. The imaginary part  of the potential comprises local terms of the form $\varphi^\dagger \delta^3(\bm{r}) \varphi$ as inherited from the four-fermion operators of NRY$_{\gamma_5}$ accounting for fermion-antifermion annihilations~\cite{Braaten:1996ix,Pineda:1998kn,Brambilla:1999xf,Brambilla:2004jw}, that we shall address explicitly in section~\ref{sec:pNRYann}.\footnote{Thermal corrections to the Coulomb potential can be calculated in pNRY$_{\gamma_5}$, by computing self-energies of the wave function field, where the typical energy and momenta in the loop diagrams are of order $M \alpha^2 $ and $T$ (\cf \myRefs\cite{Brambilla:2008cx,Brambilla:2010vq,Escobedo:2010tu,Escobedo:2008sy} for QED and QCD).} In the second line of \Eq\eqref{pNREFT_sca_0}, we see the appearance of a monopole and a quadrupole interactions as well as interactions involving the derivative in the relative distance. Such structures arise from the multipole expansion. In contrast to pNRQED and pNRQCD, the dipole term is absent. The last line of  \Eq\eqref{pNREFT_sca_0} accounts for the scalar sector, where the scalar field $\phi$ should be equally understood as being multipole expanded. The matching coefficients $d_1,...,d_5$ are inherited from NRY and they read, at tree level, as follows \cite{Biondini:2021ccr}
 \begin{equation}
     d_1=d_2=1,  \quad d_3=-\lambda, \quad  d_3=d_4=0 \, .
 \end{equation}

\begin{figure}
    \centering
    \includegraphics[width=0.95\textwidth]{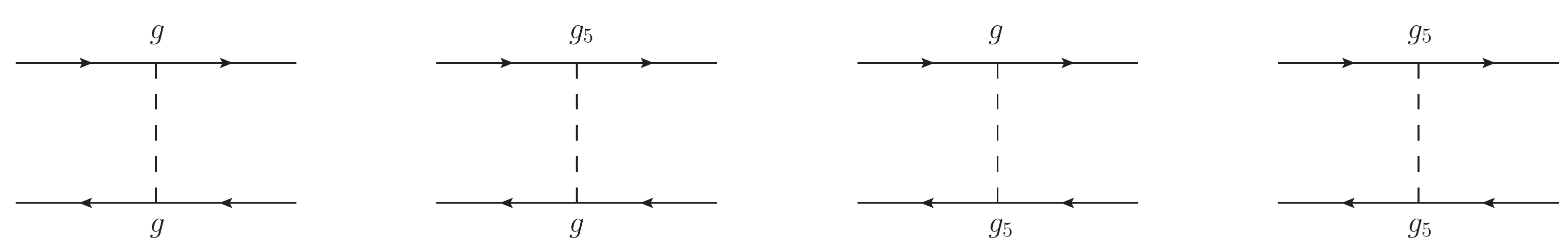}
    \caption{Leading order NRY$_{\gamma_5}$ diagrams for the potential matching with scalar and pseudo-scalar interactions required to calculate $V(\bm{p},\bm{r},\bm{\sigma}_1,\bm{\sigma}_2)$.}
    \label{fig:g_g5_pot}
\end{figure}
Let us discuss the modifications of pNRY in the presence of the pseudo-scalar coupling. There are two main aspects to be addressed. The first one concerns the modification of the potential $V(\bm{p},\bm{r},\bm{\sigma}_1,\bm{\sigma}_2)$, or simply $V$ in the following, from the inclusion of the pseudo-scalar coupling. We have 
\begin{equation}
	V(\bm{p},\bm{r},\bm{\sigma}_1,\bm{\sigma}_2) = V (\bm{p}, \bm{r},\bm{\sigma}_1,\bm{\sigma}_2) \big \vert_{ g} +
	V (\bm{p}, \bm{r},\bm{\sigma}_1,\bm{\sigma}_2) \big \vert_{ g_5},
\end{equation}
where $V (\bm{p}, \bm{r},\bm{\sigma}_1,\bm{\sigma}_2) \big \vert_{ g}$ denotes the part of the potential induced by pure
scalar interactions, while $V (\bm{p}, \bm{r},\bm{\sigma}_1,\bm{\sigma}_2) \big \vert_{ g_5}$ stems from pseudoscalar and mixed scalar-pseudoscalar contributions.

The potential is obtained by matching four-fermion off-shell Green's functions of NRY$_{\gamma_5}$ onto pNRY$_{\gamma_5}$.
The insertions of the scalar operator with the matching coefficient $d_4$ and the contributions from the 
four-fermion operators are suppressed in the power counting. This is why at the accuracy we are aiming at it is sufficient to consider tree level diagrams with two vertices involving the leading fermion-scalar and antifermion-scalar interactions
shown in figure~\ref{fig:g_g5_pot}. The first diagram in figure~\ref{fig:g_g5_pot} is identical to the one found in NRY, meaning that the leading contribution to $V(\bm{p},\bm{r},\bm{\sigma}_1,\bm{\sigma}_2)$ is the Coulomb potential $V^{(0)}=-\alpha/r$. In addition to that, one has to consider diagrams that involve pseudoscalar interactions with $g_5$, which are a distinctive feature of NRY$_{\gamma_5}$. Here one finds a mixed scalar-pseudoscalar contribution and a pure pseudoscalar term. Doing the matching in the momentum space and Fourier transforming~\cite{adkins2013} the results to the position space we obtain
\begin{align}
 V (\bm{p}, \bm{r},\bm{\sigma}_1,\bm{\sigma}_2) \big \vert_{ g_5} &= - 
 \frac{g g_5}{4 \pi M } \frac{ \bm{r} \cdot \bm{\sigma}_1-\bm{r} \cdot \bm{\sigma}_{2}}{r^3} \nonumber \\ 
 & + \frac{\pi \alpha_5}{M^2} \left[ - \frac{\bm{\sigma}_1 \cdot \bm{\sigma}_2}{3}  \delta^3 (\bm{r}) + \frac{3 \, \bm{\sigma}_1 \cdot \bm{\hat{r}} \, \bm{\sigma}_2 \cdot \bm{\hat{r}} -\bm{\sigma}_1 \cdot \bm{\sigma}_2}  {4 \pi r^3}\right]\, ,
 \label{correction_alpha_g5}
 \end{align}
where $\bm{\hat{r}} = \bm{r}/r$ with $\bm{\sigma}_1$ ($\bm{\sigma}_2$) being the spin matrix of the two-component particle (antiparticle) field in the fermion bilinear. In order to estimate the relative importance of such corrections, we set $1/r \sim M \alpha$. Therefore, the mixed and pure-pseudoscalar terms are of order $M \alpha^3 (g_5/g)$  and $ M \alpha^4 (g_5^2/g^2)$ respectively. 
 As for the pure pseudoscalar potential, our result agrees with earlier derivations~\cite{Daido:2017hsl,Kahlhoefer:2017umn}. In the case of $g_5 \approx g$, the pseudo-scalar exchange induces an $\alpha$- and $\alpha^2$-suppressed contributions as compared to the leading scalar Coulomb potential. In the following we assume the pseduoscalar coupling $g_5$ to be smaller than $g$, which renders the contributions from \Eq\eqref{correction_alpha_g5} even more suppressed. Hence, we work with the leading Coulombic energy levels and the Bohr radius given by
\begin{equation}
E_n = -\frac{M \alpha^2}{4 n^2} = -\frac{1}{M a_0^2 n^2}\, , \quad a_0 \equiv \frac{2}{M \alpha}\, .
    \label{Coulomb_en_levels}
\end{equation}
The second aspect to be discussed is the modification of the second line in \Eq\eqref{pNREFT_sca_0}, namely the interaction between fermion-antifermion pairs and ultra-soft pseudoscalars. One may compute the corresponding vertices by multipole expanding terms containing $g_5$ in \Eq\eqref{bilinear_psi_5} and \Eq\eqref{bilinear_chi_5}, where we refer to
appendix~\ref{sec:appendix-pnry} for further details.
The main outcome is that the pseudoscalar induced ultra-soft transitions are suppressed by powers of $\alpha$ and $g_5/g$ with respect to the ultra-soft contributions displayed in the second line of \Eq\eqref{pNREFT_sca_0}. 
  
In summary, upon neglecting the suppressed corrections from the pseudo-scalar coupling and assuming the hierarchy of scales given in \Eq\eqref{hiearchy}, we find that pNRY$_{\gamma_5}$ is formally identical to the in-vacuum case already studied in \myRef\cite{Biondini:2021ccr}. However, there is a crucial difference: the vertices and propagators now have to be understood in a finite temperature field theory, since the EFT in \Eq\eqref{pNREFT_sca_0} comprises the temperature as a dynamical scale. 

We conclude the section by stressing that a different choice of the coupling arrangement, namely $g_5 > g$, would change the low-energy theory \Eq\eqref{pNREFT_sca_0} in a number of ways. First, one would need to include the pseudoscalar and mixed scalar-pseudoscalar contributions to the potential given in \Eq\eqref{correction_alpha_g5}. In this case the leading order potential $V$ is not Coulombic anymore, and nice analytic results for the quantum mechanical matrix elements should be superseded by a purely numerical evaluation. Second, one has to include the ultra-soft contribution originated by the pseudoscalar vertex as well.	Phenomenologically, we expect such a regime to be important when $g_5/g \sim 1/\alpha$, that lifts the suppression we rely on.

 \subsection{Thermal propagators}
 \label{sec:thermal_prop}
Let us introduce the finite temperature formalism needed for the calculation of the bound-state formation and dissociation rate in a thermal environment. In particular, we shall provide the Feynman rules for heavy fermion-antifermion pairs and light scalars of the Yukawa model as given in \Eq\eqref{lag_mod_0}. We work within the real-time formalism (RTF) of thermal field theory. Real-time expectation values depend on how the contour of the time integration in the partition function is deformed to include real times. The modified contour has two lines stretching along the real-time axis. 
The main practical consequence of this is that the number of degrees of freedom doubles. Fields on the upper branch enjoy the usual time ordering, whereas for the fields living on the lower branch the time ordering is reversed. The physical degrees of
freedom describing initial and final states, are of type 1 (upper branch), while fields of type 2 stem from the lower branch. Propagators are represented by $2 \times 2$ matrices, and they can mix fields of type 1 with fields of type 2. The vertices, however, do not couple fields of different types. For further details, we refer to the standard textbook treatment~\cite{Bellac:2011kqa,Kapusta:1989tk} of the subject. 

In order to introduce the finite temperature propagator of a darkonium system, we can build upon its QCD counterpart, the heavy quarkonium~\cite{Brambilla:2008cx,Brambilla:2010vq}. The two systems share many similarities, since both are made of  a heavy particle-antiparticle pair with a mass much larger than the temperature of the thermal bath. The non-relativistic propagator of a fermion-antifermion pair interacting
through a potential $V(r)$ reads~\cite{Brambilla:2008cx} \footnote{In heavy-ion collisions the heavy quarkonium pairs are taken to be far away from chemical and kinetic equilibration. In the DM case, it is possible to assume kinetic equilibration after the chemical decoupling. In any case, the condition $M \gg \pi T$ holds for the problem at hand, and the thermal contribution to the heavy-pair propagator, here the darkonium, is exponentially suppressed as well.}
\begin{equation}
    \Delta_{\varphi}(k_0,k)=\left( 
    \begin{array}{cc}
        \frac{i}{k_0 - h+ i \eta} & 0 \\
       2\pi \delta(k_0 - h) & \frac{-i}{k_0 - h - i \eta}
    \end{array}
    \right)
\label{Prop_pair_RTF}
\end{equation}
where the Hamiltonian $h$ at leading order comprises   $M \alpha^2$ terms, namely $h^{(0)}=k^2/M + V(r)$. As first noted in \myRef\cite{Brambilla:2008cx}, since the  $[\Delta_{\varphi}(k_0,k)]_{12}$ component vanishes,  the fermion-antifermion field of type 2 never appears in amplitudes with final and initial states being physical fields of type 1. Hence, it comes as a great simplification in the non-relativistic theories at finite temperature  to discard the type-2 fields when considering physical amplitudes. Notice that even though in RTF the heavy-pair propagator acquires a matrix form, it does not depend on the temperature.

Next, one needs the thermal propagator of the scalar mediator. In this work we assume the in-vacuum and thermal masses of the scalar to be negligible with respect to the binding energy. As discussed earlier, it suffices to assume a very small scalar self-coupling $\lambda$ in order not to generate a sizable $m_T$. The $2 \times 2$ free scalar propagator at finite temperature reads
\begin{align}
    \Delta_{\phi} (k_0,k)&=\left( 
    \begin{array}{cc}
        \frac{i}{k_0^2 - k^2 + i \eta} & \theta(-k_0) \,  2\pi \delta(k_0^2 - k^2) \\
       \theta(-k_0)  \, 2\pi \delta(k_0^2 - k^2) & \frac{-i}{k_0^2 - k^2 - i \eta}
    \end{array}
    \right) 
    \nonumber 
\\
& + 2 \pi \delta(k_0^2 - k^2) n_\textrm{B}(|k_0|) \left(  \begin{array}{cc} 
    1 \enspace & 1\\
    1 \enspace & 1
    \end{array} \right) \, ,
\label{Prop_scalar_RTF}
\end{align}
where $n_{\textrm{B}}(x)=1/(e^{x/T}-1)$ is the Bose-Einstein equilibrium distribution. 
Due to the decoupling of the type-2 heavy pair fields, in the following we will only need the 11-component of the heavy fermion-antifermion propagators. Accordingly, the relevant scalar propagator is given by $[\Delta_{\phi} (k_0,k)]_{11}$, whereas the other entries of $[\Delta_{\phi} (k_0,k)]$ are irrelevant at the order we are working in this paper. One-loop corrections to the scalar propagator, where all entries of $[\Delta_{\phi} (k_0,k)]$ contribute, as well as the connection to additional physical processes between the heavy pairs and the thermal environment are deferred to another study in preparation~\cite{Biondini_prep}. 

\section{Thermal cross section and thermal width}
\label{sec:pNRY_application}
We are now in the position to carry out the calculation of the bound-state formation cross section in pNRY$_{\gamma_5}$ at finite temperature, thus extending the result presented in \myRef\cite{Biondini:2021ccr} for the in-vacuum case. Moreover, we compute the inverse process, namely bound-state dissociation, which is a genuine thermal effect. Let us stress that as far as the extraction of the bound-state formation and dissociation are concerned, when working at the accuracy we are aiming at in this work, pNRY and pNRY$_{\gamma_5}$ are equivalent. This is not true when dealing with heavy-pair annihilations and bound-state decays, as will be shown in section~\ref{sec:pNRYann}. 

\subsection{Bound-state formation via scalar emission}
\label{sec:bsf}
\begin{figure}[t!]
    \centering
    \includegraphics[width=0.95\textwidth]{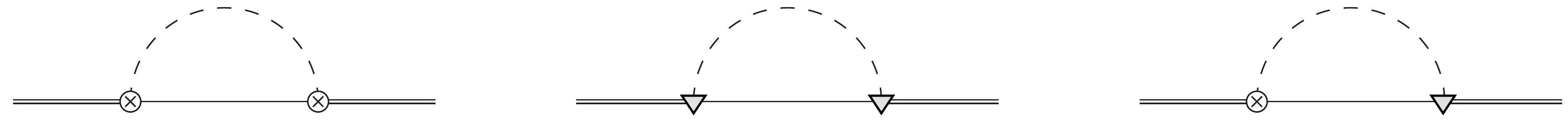}
    \caption{Self-energy of a heavy pair in a scattering state (solid double line) induced by the quadrupole (crossed vertex) interaction, derivative interaction (triangle vertex) and their mixing. The internal single line stands for a pair in a bound state, the dashed line is the scalar mediator.}
    \label{fig:pNRY_cross_sec}
\end{figure}

 The main advantage in addressing bound-state calculations within pNRY$_{\gamma_5}$ is that we can
 describe the system using appropriate degrees of freedom at the energy scale of interest, which are heavy fermion-antifermion pairs as well as ultrasoft and thermal scalars. The bound-state formation cross section can be extracted from the imaginary part of the self-energy diagram of the pair, that reads 
\begin{equation}
     \sigma_{\textrm{bsf}}v_{\textrm{rel}} = \braket{ \bm{p } | {2\rm{Im}}(-\Sigma_s) | \bm{p} } \, ,
     \label{master_cross}
     \end{equation}
 where $\ket{\bm{p}}$ denotes a scattering state.
     The self-energy diagrams contributing to $\Sigma_s$ are shown in figure~\ref{fig:pNRY_cross_sec}. 

     Out of the three vertices in the second line of \Eq\eqref{pNREFT_sca_0}, only the quadrupole and the derivative interactions contribute. As has already been observed in the literature~\cite{Wise:2014jva,Petraki:2016cnz}, the monopole contribution is zero due to the vanishing overlap between the scattering and bound state wave functions. This still holds at finite temperature. The momentum region in the loop diagrams are those still dynamical in pNRY$_{\gamma_5}$ according to the assumed hierarchy of scales, namely the ultrasoft scale and the temperature scale, which are taken to be of the same order. 
     
     As an example, let us explicitly show the contribution to the self-energy as originating from the quadrupole interaction, and elucidate on the main steps towards the thermal cross section. We adopt dimensional regularization (DR) with $D= 4 - 2 \epsilon$. The corresponding self-energy, leftmost diagram figure~\ref{fig:pNRY_cross_sec}, reads
  \begin{align}
     \Sigma^{\mathcal{Q}} &= -i \frac{\pi \alpha}{4} \mu^{4-D} r^i r^j \, \int \frac{d^D k}{(2 \pi)^D} \frac{i}{P^0-h-k^0+i\eta}  \, k^i k^j k^m k^n  \nonumber 
     \\
     & \times \left[ \frac{i \,}{k_0^2-k^2 +i\eta}  + 2 \pi \delta (k_0^2-k^2) n_{\textrm{B}} (|k_0|)\right] \, r^m r^n \, ,
     \label{self_energy_quad}
 \end{align}
 where one may notice the appearance of powers of the scalar three momenta in  \eqref{self_energy_quad} as induced by the action of the derivative operator $\nabla_{\bm{R}}$ on the scalar propagator. Notice that here $\bm{r}$ is a quantum mechanical operator that generates expectation values when acting on the scattering states. 
  The scalar propagator contains both an in-vacuum and a finite temperature contribution, that are manifestly disentangled in RTF and at the order we perform the calculation. Next, we insert a complete set of bound states, thus  ensuring  that the internal propagator in the loop diagram describes the propagation of the discrete states of the particle-antiparticle spectrum
 \begin{equation}
	\frac{i}{P^0-h-k^0+i\eta} = \sum_n \frac{i}{P^0-h-k^0+i\eta} \ket{n} \bra{n}  =
	\sum_n \frac{i}{E_p-E_n-k^0+i\eta}  \ket{n} \bra{n},
\end{equation}
with 
 \begin{equation}
     E_p-E_n \equiv \Delta E^p_n = \frac{p^2}{M} + \frac{M \alpha^2}{4 n^2} \, ,
     \label{diff_scat_bound}
 \end{equation}
 at leading order, where $p = M v_{\textrm{rel}}/2$ is the relative momentum  of the pair in a scattering state. Finally, one can extract the imaginary part of the in-vacuum contribution with the standard cutting rules, whereas for the term involving the finite temperature contribution to the scalar propagator, which already gives an on-shell thermal distribution of propagating scalar fields, we simply select the imaginary part of the bound-state propagators with the relation $1/(x \pm i \eta)= \textrm{P.V.}(1/x) \mp i \delta(x)$ (\cf\cite{Brambilla:2008cx} for a similar derivation in pNRQCD), where $\textrm{P.V.}$ stands for the Cauchy principal value. The result for the quadrupole-induced cross section reads
\begin{eqnarray}
     \sigma_{\textrm{bsf}}^{\mathcal{Q}}v_{\textrm{rel}} \big \vert_{T} = \frac{\alpha}{120}    \sum_{n} (\Delta E_n^p)^5\left[ |\langle\bm{p} | \bm{r}^2 |n  \rangle|^2 + 2|\langle\bm{p} | r^i r^j |n  \rangle|^2\right] \left[ 1+ n_{\textrm{B}}(\Delta E^p_n) \right]\, . 
     \label{cross_section_quadrupole_T}
\end{eqnarray}
This expression contains non-trivial quantum mechanical expectation values that must be explicitly evaluated 
in order to arrive at a final result. 
The thermal cross section \eqref{cross_section_quadrupole_T} is a generalization of the in-vacuum counterpart derived in \myRef\cite{Biondini:2021ccr} and constitutes an original result of the present work. It is easy to see that for $T \to 0$
we readily recover the in-vacuum result.

The contributions from the derivative vertex (middle diagram in figure~\ref{fig:pNRY_cross_sec}) and the mixed diagrams (rightmost diagram in figure~\ref{fig:pNRY_cross_sec}), that comprises one quadrupole and one derivative vertices, have to be added as well. These contribute at the same order in the power counting~\cite{Biondini:2021ccr}. The finite temperature cross section, upon including all the relevant diagrams reads
\begin{eqnarray}
 \sigma_{\textrm{bsf}} v_{\textrm{rel}} \big \vert_{T} =  \sigma_{\textrm{bsf}} v_{\textrm{rel}} \big \vert_{T=0}    \left[ 1+ n_{\textrm{B}}(\Delta E^p_n) \right] \, ,
 \label{bsf_matrix_element_T}
\end{eqnarray}
where the in-vacuum cross section is
\begin{eqnarray}
\sigma_{\textrm{bsf}} v_{\textrm{rel}} \big \vert_{T=0} &=&  \frac{\alpha}{120 }    \sum_{n} (\Delta E_n^p) ^5 \left[ |\langle\bm{p} | \bm{r}^2 |n  \rangle|^2 + 2 |\langle\bm{p} | r^i r^j |n  \rangle|^2\right]  \nonumber
 \\
 && + 2 \alpha  \sum_{n} \Delta E_n^p  \, \,  \Big \vert \Big \langle \bm{p} \Big \vert \frac{\nabla^2_{\bm{r}}}{M^2} \Big \vert n  \Big \rangle \Big \vert^{2}  \, 
 \nonumber
 \\
 &&-  \frac{\alpha}{3}  \sum_{n} (\Delta E_n^p)^3 \, \textrm{Re} \left[ \Big \langle \bm{p} \Big \vert \frac{\nabla^2_{\bm{r}}}{M^2} \Big \vert n  \Big \rangle  \langle n | \bm{r}^2 | p \rangle \right]  \, .
 \label{bsf_matrix_element_T0}
\end{eqnarray}
As expected from the quadrupole-induced thermal cross section in \Eq\eqref{cross_section_quadrupole_T}, $\left[ 1+ n_{\textrm{B}}(\Delta E^p_n) \right]$ factors out in each term. The second line in \Eq\eqref{bsf_matrix_element_T0} corresponds to the derivative contribution, whereas the third line accounts for the mixed contribution from the quadrupole-derivative interaction. Let us stress that the result given in \Eq\eqref{bsf_matrix_element_T} is obtained without making use of any semi-classical construction from Boltzmann equations to describe particle collisions in the plasma. The thermal character of the cross section is rigorously derived from a quantum field theory at finite temperature, here pNRY$_{\gamma_5}$. One may recognize the typical factor that one would introduce according to the Boltzmann prescription to accompany the produced scalar particle $\phi$ in the final state. There is no such factor for the heavy darkonium, since in our setting $M \gg T$. This means that the corresponding Bose distribution is exponentially suppressed and cannot show up in our EFT. Another relevant derivation that exploits a pNREFT can be found in \myRef\cite{Binder:2020efn}, where a carbon copy of QED for the dark sector is employed to derive a thermal cross section. However, the identification of the thermal cross sections appears to be partly based on the network of Boltzmann equations. 

\begin{figure}[t!]
    \centering
    \includegraphics[width=0.45\textwidth]{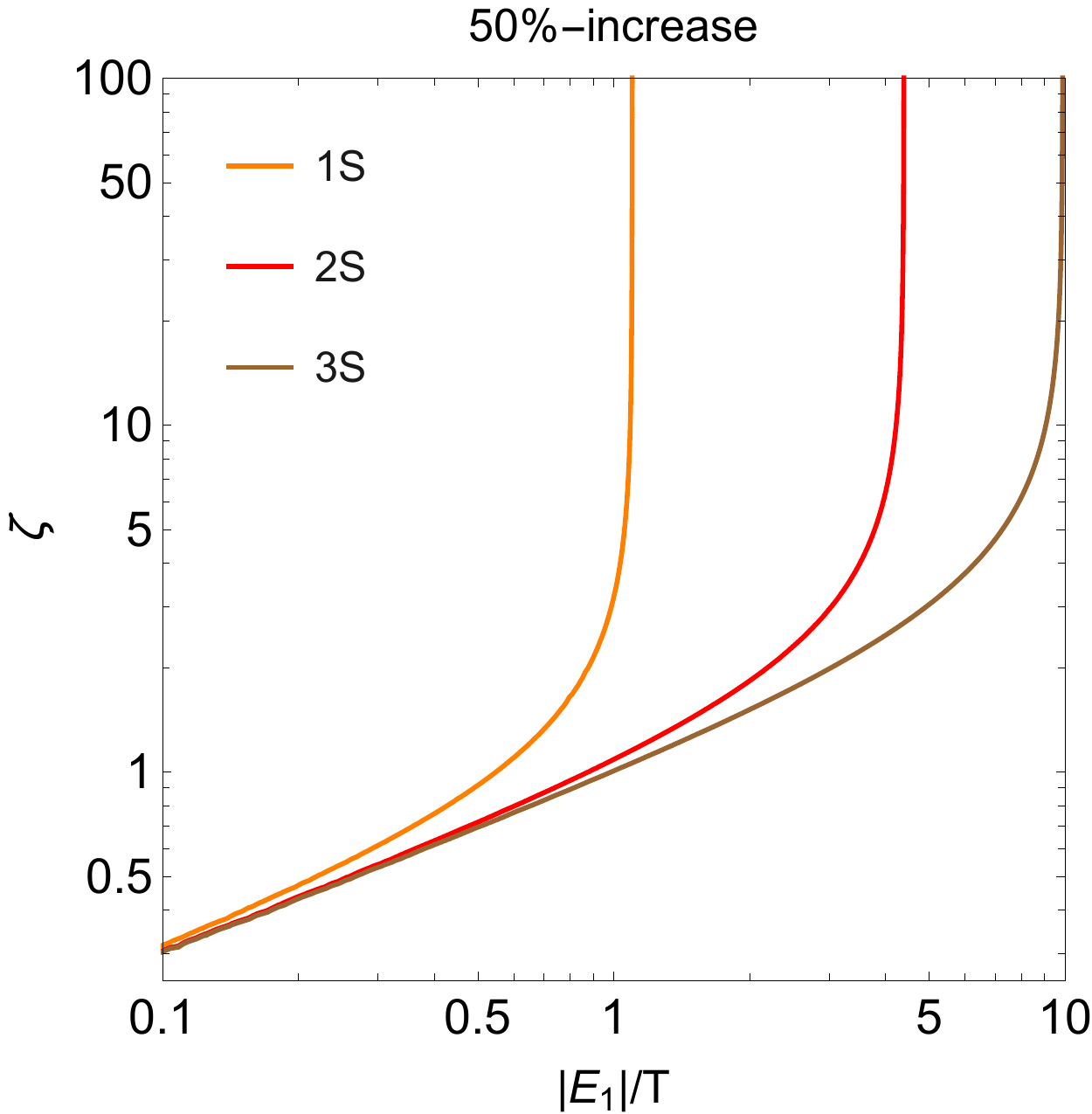}
    \hspace{0.35 cm}
        \includegraphics[width=0.45\textwidth]{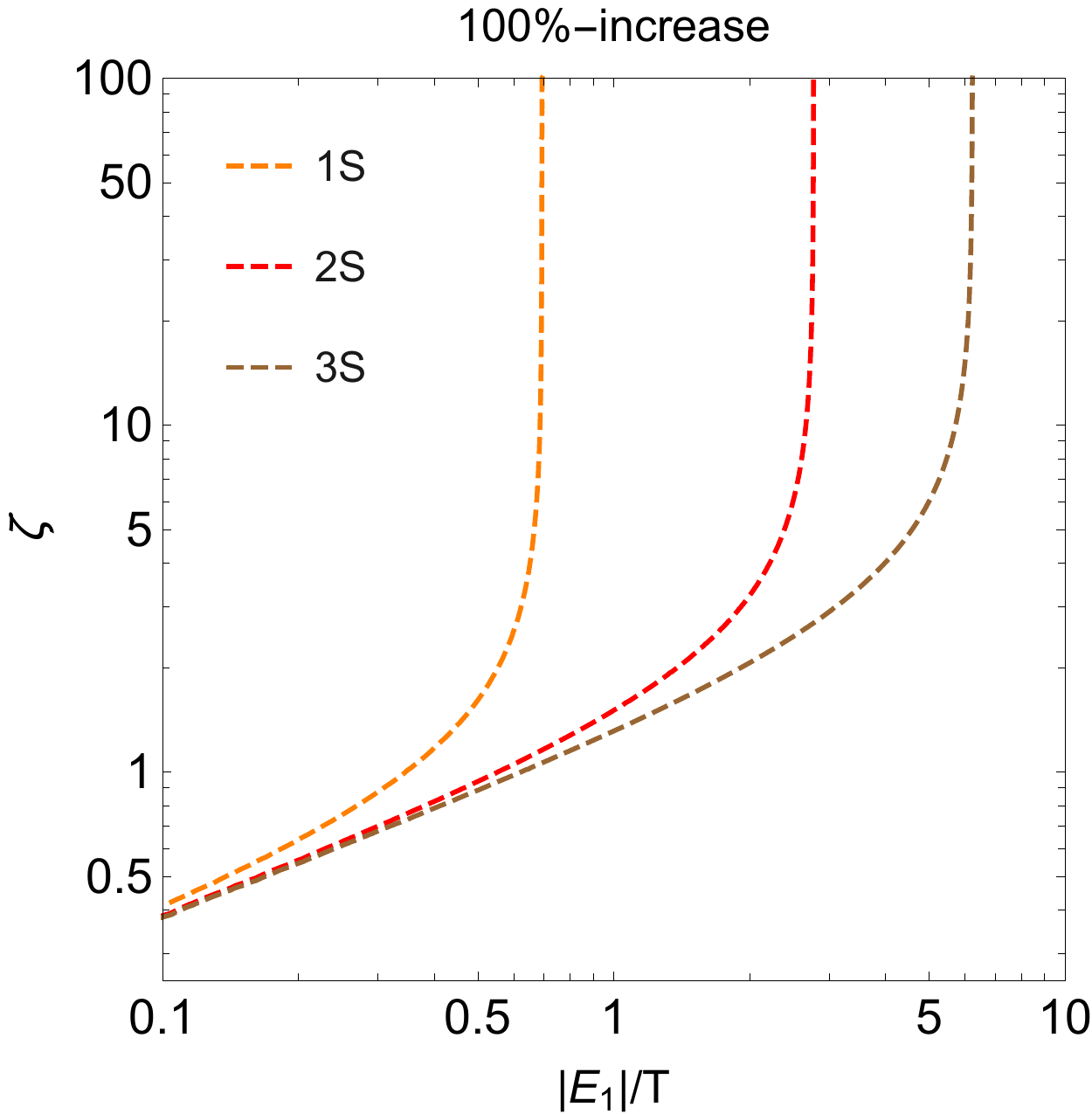}
    \caption{Contour levels for the ratio of the thermal and in-vacuum cross sections $1+ n_{\textrm{B}}(\Delta E^p_n)$ in the $(|E_1|/T, \zeta)$-plane. The solid (dashed) lines stand for a 50\% (100\%) increase of the in-vacuum cross section due to the Bose enhancement.}
    \label{fig:plot_1}
\end{figure}

The cross section \eqref{bsf_matrix_element_T} can give us a first indication of the importance of thermal effects in the temperature region allowed for the hierarchy of scales of interest. The factorization of the in-vacuum and thermal parts of the cross section allows to rewrite the ratio $ \sigma_{\textrm{bsf}} v_{\textrm{rel}} \big \vert_{T}/\sigma_{\textrm{bsf}} v_{\textrm{rel}} \big \vert_{T=0}$ as $\left[ 1+ n_{\textrm{B}}(\Delta E^p_n) \right] $, and we express the argument of the Bose distribution as $ |E_1|/T(1/\zeta^2+1/n^2)$. Here we defined $\zeta =\alpha /v_{\textrm{rel}}$, which is commonly used in the literature to estimate the relative velocity of the particles in the pair in a scattering state, while $E_1=-M \alpha^2/4$ stands for the ground-state energy. In figure~\ref{fig:plot_1}, we show the contour lines, solid and dashed respectively, for a 50\% and 100\% increase of the in-vacuum cross section due to thermal effects. One may notice  that for sufficiently large value of $\zeta$ or sufficiently small relative velocities for a fixed $\alpha$, the argument of the Bose distribution becomes independent of $\zeta$, and this is reflected in the solid and dashed curves reaching an asymptotic behavior. As we adopt the variable $|E_1|/T$ for the different bound states, the same increase of the in-vacuum cross sections for the excited states occurs at smaller temperature. This essentially implies that $|E_n|<|E_{n'}|$ for $n > n'$. A larger increase of the cross section requires larger temperatures, which corresponds to smaller values of $|E_1|/T$. 

The result in terms of pNRY matrix elements allows us to address the calculation of excited states as well. We provide  explicit expressions for the ground state and the excited  $2S$ and $3S$ states (\cf appendix~\ref{sec:appendix-me} for details). The corresponding in-vacuum cross sections read
\begin{subequations}
\begin{align}
    \sigma_{\textrm{bsf}}^{1S} v_{\textrm{rel}} \big \vert_{T=0} &= \frac{\pi \alpha^4}{M^2} S(\zeta)  \frac{2^6}{15}\frac{\zeta^2 (7 + 3 \zeta^2)}{(1+\zeta^2)^2} e^{-4 \zeta \arccot(\zeta)} \, ,
    \label{bsf_1S}
    \\
     \sigma_{\textrm{bsf}}^{2S} v_{\textrm{rel}} \big \vert_{T=0} &= \frac{\pi \alpha^4}{M^2}  S(\zeta) \frac{2^5}{15} \frac{\zeta^2 (448 + 528 \zeta^2 + 100 \zeta^4 +15  \zeta^6)}{(4+\zeta^2)^4} e^{-4 \zeta \arccot(\zeta/2)} \, , 
     \label{bsf_2S}
     \\
      \sigma_{\textrm{bsf}}^{3S} v_{\textrm{rel}} \big \vert_{T=0} &=  \frac{\pi \alpha^4}{M^2}  S(\zeta) \frac{2^6}{135}  e^{-4 \zeta \arccot(\zeta/3)} \quad
      \nonumber\\
      & \times \frac{\zeta^2 (137781 + 186624 \zeta^2 + 57618 \zeta^4 + 8376 \zeta^6 + 881 \zeta^8)}{(9+\zeta^2)^5} \, ,
      \label{bsf_3S}
\end{align}
\end{subequations}
where the Coulombic $S$-wave Sommerfeld factor is defined as $S(\zeta)= 2\pi \zeta/(1-e^{-2 \pi \zeta})$. 
The result for the ground state in \Eq\eqref{bsf_1S} has been adopted from \myRef\cite{Biondini:2021ccr}, whereas the expressions for the excited states are an original contribution of this work. The in-vacuum cross sections are consistent with the unitarity bound on the annihilation cross section in the non-relativistic regime~\cite{Griest:1989wd}, and have been lately revisited and extended to the bound-state formation process~\cite{vonHarling:2014kha,Petraki:2015hla,Baldes:2017gzw}. The velocity scaling at $\zeta \gg 1$ obeys $\sigma_{\textrm{bsf}} \propto  1/v_{\textrm{rel}}^2$, which equally holds for our findings in \Eqs\eqref{bsf_1S}--\eqref{bsf_3S}. 
\begin{figure}[t!]
    \centering
    \includegraphics[width=0.45\textwidth]{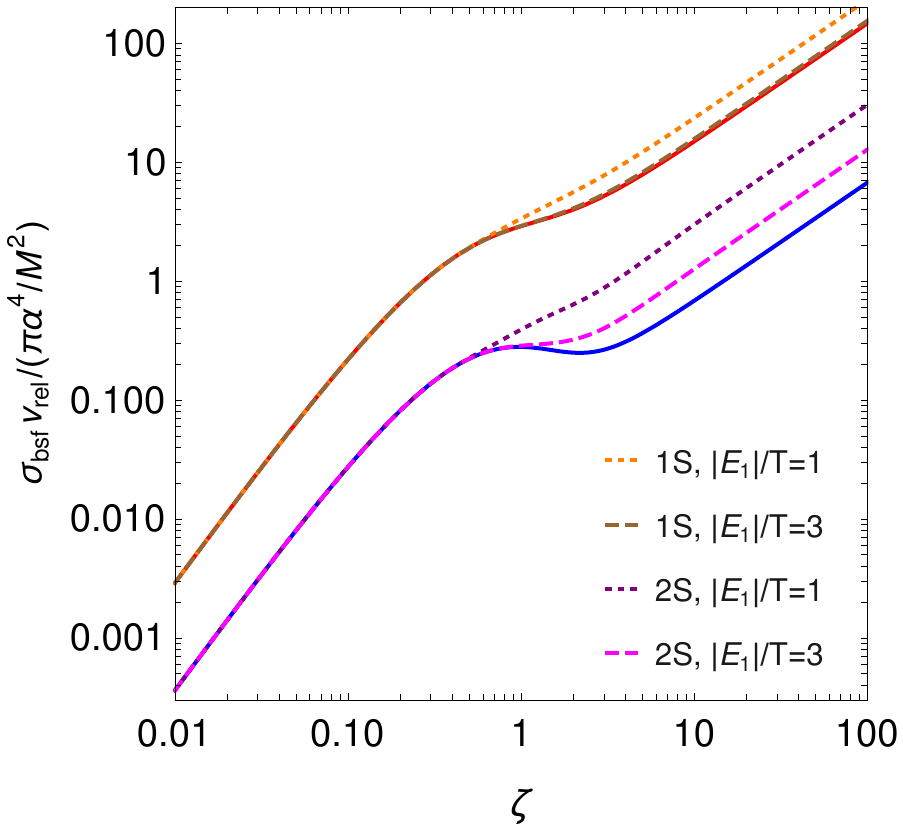}
    \hspace{0.5 cm}
     \includegraphics[width=0.45\textwidth]{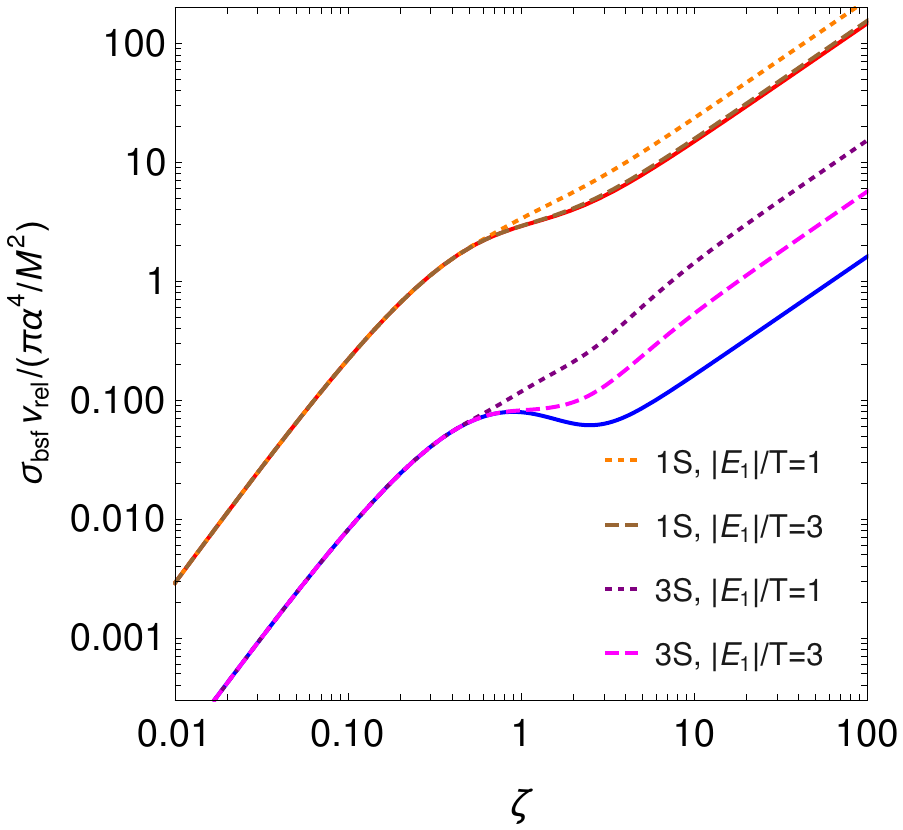}
    \caption{Bound-state formation cross section, normalized by the factor $\pi \alpha^4/M^2$, for the $1S$ and $2S$ states as function of $\zeta$ (left plot). Two choices for $|E_1|/T$ are considered. The comparison between the $1S$ and $3S$ states is shown in the right plot.}
    \label{fig:fig2}
\end{figure}

Figure~\ref{fig:fig2} shows the in-vacuum (solid lines) and thermal (dotted and dashed lines) cross sections for the $1S$ and $2S$ states, plotted as functions of $\zeta$ for two different values of $|E_1|/T$. Some comments are in order. First, the in-vacuum and thermal cross sections are virtually identical for sufficiently small values of $\zeta$, whereas the thermal cross sections get larger than the vacuum ones for $\zeta >1$. Second, the larger the temperature, \ie the smaller the values of $|E_1|/T$, the sooner $\sigma_{\textrm{bsf}}v_{\textrm{rel}}|_{T}$ gets enhanced by the finite-temperature nature of the emitted scalar particle. For $|E_1|/T=3$, the in-vacuum and thermal cross section for the ground-state are the same, whereas for $\zeta \simg 10$ there is still a factor of 2 or 4 in the case of the excited state $2S$ or $3S$ respectively.

\begin{figure}[t!]
    \centering
    \includegraphics[width=0.45\textwidth]{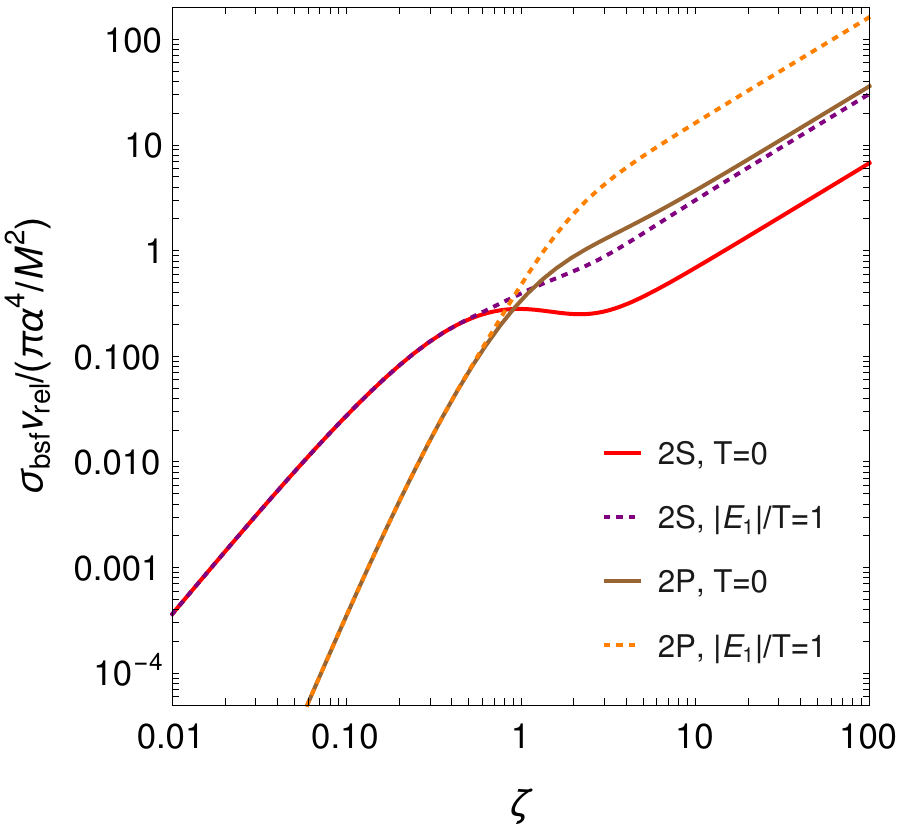}
    \hspace{0.5 cm}
     \includegraphics[width=0.45\textwidth]{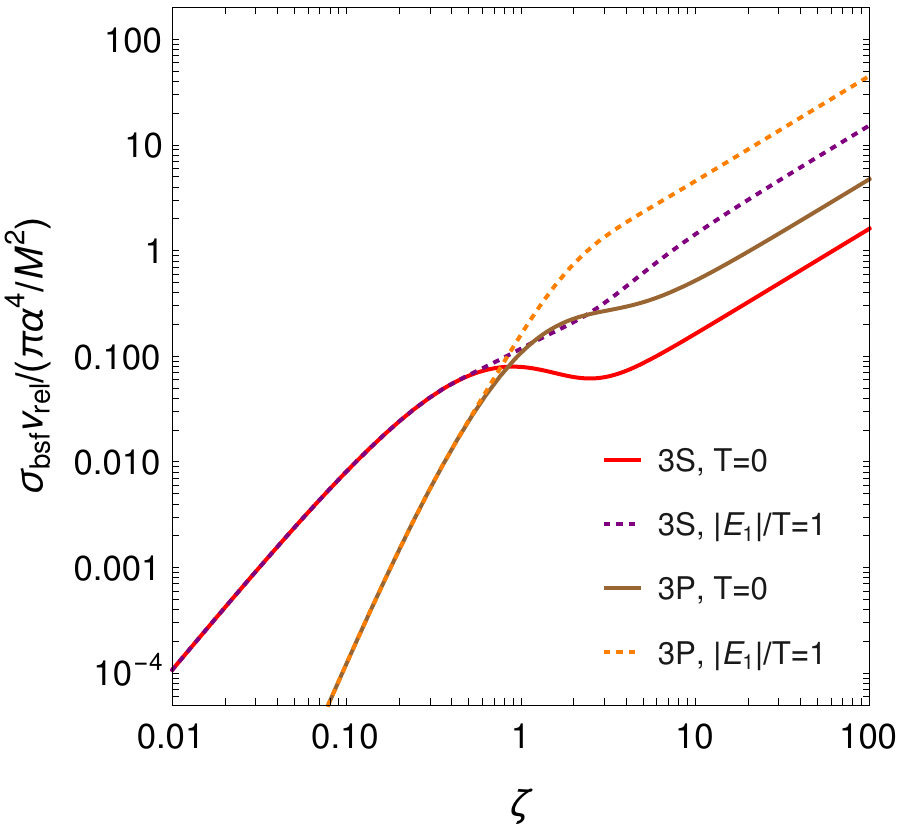}
    \caption{Bound-state formation cross section for the $\ket{2P}$ and $\ket{3P}$ states as a function of $\zeta$. The corresponding bound-state formation for the $\ket{2S}$ and $\ket{3S}$ states are shown. The in-vacuum and thermal cases, for $|E_1|/T=1$ are considered.}
    \label{fig:fig4}
\end{figure}

In addition to the bound-state formation in $nS$ states, we can easily compute the formation in $P$-waves. The result for the $\ket{21m}$ and $\ket{31m}$ states, where we sum over the magnetic component of the angular momentum read
\begin{align}
    \sum_m \sigma_{\textrm{bsf}}^{21m} v_{\textrm{rel}} \big \vert_{T=0} &= \frac{\pi \alpha^4}{M^2}  S(\zeta) \frac{2^9}{15} \frac{\zeta^4 (1+\zeta^2)(36+5 \zeta^2)}{(4+\zeta^2)^4} e^{-4 \zeta \arccot(\zeta/2)} \, , 
     \label{bsf_2P}
     \\
    \sum_m  \sigma_{\textrm{bsf}}^{31m} v_{\textrm{rel}} \big \vert_{T=0} &= \frac{\pi \alpha^4}{M^2}  S(\zeta) \frac{3 \cdot 2^{11} }{5} \frac{\zeta^4 (1+\zeta^2)(729+270 \zeta^2+26 \zeta^4 +\zeta^6)}{(9+\zeta^2)^6} e^{-4 \zeta \arccot(\zeta/3)}   \,.
      \label{bsf_3P}
\end{align}
We show the corresponding cross sections in figure~\ref{fig:fig4}, where they are plotted together with the corresponding bound-state formation cross section for $2S$ and $3S$ states. One may notice that for $\zeta \simg 1$ the cross section is larger for the  $nP$ states. Moreover, we have $\ket{2P}/\ket{2S} \approx 5.3$ and $\ket{3P}/\ket{3S} \approx 3.1$ for $\zeta \simg 10$ where the ratios stay constant.\footnote{Notice that the combination $S_p \equiv S(\zeta)(1+\zeta^2)$ appearing in \Eqs\eqref{bsf_2P} and \eqref{bsf_3P} defines the $P$-wave Sommerfeld factor. We shall explicitly see how this quantity arises from the scattering-state wave function at the origin in section~\ref{sec:pNRYann}. } Also \Eqs \eqref{bsf_2P} and \eqref{bsf_3P} respect the unitarity bound for large $\zeta$ values.

\begin{figure}[t!]
    \centering
    \includegraphics[width=0.45\textwidth]{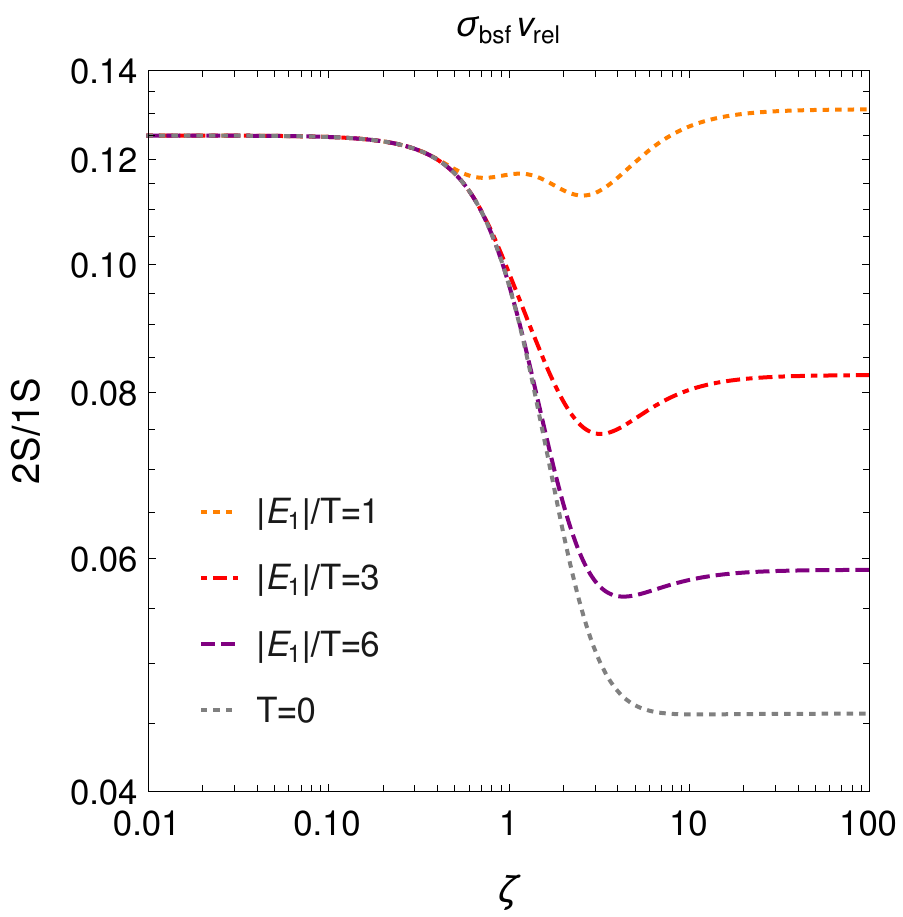}
    \hspace{0.5 cm}
     \includegraphics[width=0.45\textwidth]{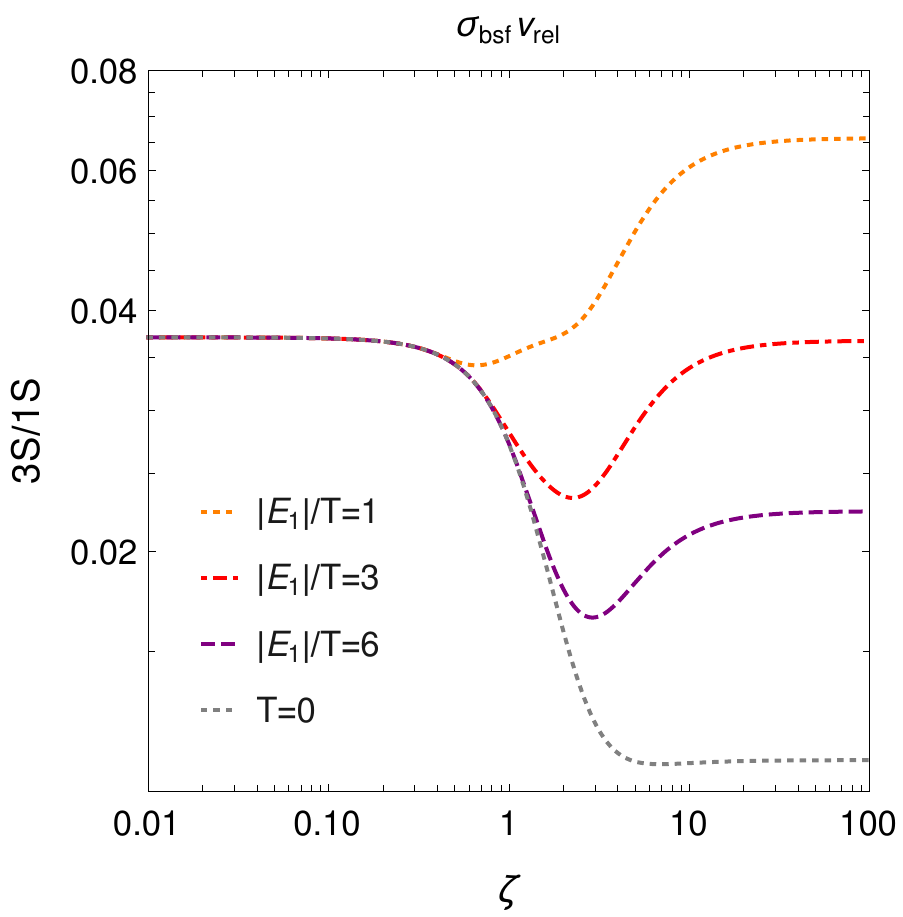}
    \caption{The ratio $\sigma^{nS}_{\textrm{bsf}}v_{\textrm{rel}}/\sigma^{1S}_{\textrm{bsf}}v_{\textrm{rel}}$ for $n=2,3$ is shown for both the in-vacuum (gray-dotted line) and thermal cross sections (colored curves).}
    \label{fig:fig3}
\end{figure}

We close the section by inspecting more closely the thermal effects in the ratios of the bound-state formation cross section $\sigma_{\textrm{bsf}}v_{\textrm{rel}}$ for the different $nS$ states. In particular, we consider $\sigma^{nS}_{\textrm{bsf}}v_{\textrm{rel}}/\sigma^{1S}_{\textrm{bsf}}v_{\textrm{rel}}$ for $n=2,3$. In figure~\ref{fig:fig3}, we show the ratio of the in-vacuum bound state formation cross sections (gray dotted line), as well as the corresponding ratio when thermal cross sections are considered, the latter for some benchmark values of $|E_1|/T$. For $\zeta \simg 1$ the ratio of the thermal bound-state formation cross section deviates from the in-vacuum case. As for the comparison between the $1S$ and $2S$ states (\cf left panel of figure~\ref{fig:fig3}), one may see how the ratio is almost constant in the whole range of $\zeta$ for $|E_1|/T=1$, and there is no sudden decrease for large $\zeta$-values (dotted yellow line contrasted with dotted gray). In the case of the $3S$ state for the same choice $|E_1|/T=1$, the ratio apparently increases with $\zeta$ in the region $\zeta \simg 1$. We note in passing, that within pNRY$_{\gamma_5}$ the relative velocity of the scattering states is typically $v_{\textrm{rel}} \siml \alpha$ (hence $\zeta \simg 1$). Indeed, 
in this regime the relevant degrees of freedom are non-relativistic fermions with momenta of order $M \alpha$ and low-energetic scalars.

\subsection{Bound-state dissociation}
\label{sec:bsd}
\begin{figure}[t!]
    \centering
    \includegraphics[width=0.95\textwidth]{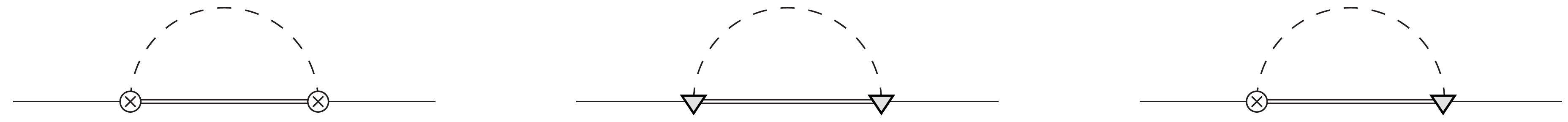}
    \caption{Self-energy of a heavy pair in a bound state (solid line) induced by the quadrupole (crossed vertex) interaction, derivative interaction (triangle vertex) and their mixing. The internal single line stands for a heavy pair in a scattering state.}
    \label{fig:pNRY_width_sec}
\end{figure}

In the following we shall take pNRY$_{\gamma_5}$ as a starting point for the computation of the dissociation rate of a bound-state, as triggered by the absorption of a scalar from the thermal medium. As it was done for the bound-state formation, we again consider the self-energy of the heavy pair and exploit the optical theorem.
The relevant diagrams are shown in figure~\ref{fig:pNRY_width_sec}, where one may see that this time the external wave function field describes a bound state, rather than a scattering state. The optical theorem provides the master formula that we use to extract the thermal width as follows
\begin{equation}
    \Gamma^n_{\textrm{bsd}}     =\langle n | {2\rm{Im}}(-\Sigma_b) | n \rangle  \, ,
    \label{master_width}
\end{equation}
where the subscript ``bsd'' stands for bound-state dissociation. In order to illustrate the main steps as it was done for the bound-state formation, we again consider  the contribution that originates from the leftmost diagram of figure~\ref{fig:pNRY_width_sec}, which contains two quadrupole vertices. Formally, the self-energy of the bound state is identical to $\Sigma^{\mathcal{Q}}$ in \Eq\eqref{self_energy_quad}. However, the incoming state has an overall negative energy, and one has to insert a complete set of \emph{scattering states} with positive energy as in
\begin{align}
	\frac{i}{P^0-h-k^0+i\eta} &= \int \frac{d^3 \bm{p}}{(2 \pi)^3} \frac{i}{P^0-h-k^0+i\eta} \ket{\bm{p}} \bra{\bm{p}} \nonumber\\
	 &= \int \frac{d^3 \bm{p}}{(2 \pi)^3} \frac{i}{E_n-E_p-k^0+i\eta}  \ket{\bm{p}} \bra{\bm{p}}  \, .
\end{align}
This readily explains the difference in the normalization and mass dimension of the bound and scattering state wave functions, that results in obtaining a width for the former and a cross section for the latter from \Eq\eqref{master_width} and \Eq\eqref{master_cross} respectively. 
Then, the second and most relevant difference arises when extracting the imaginary part. As the sign of the energy difference has changed, the real process at $T=0$ (first term in the scalar propagator $[\Delta_{\phi}(k_0,k)]_{11}$)
becomes kinematically forbidden, and only the finite temperature contribution is left. This conforms with the scalar-induced dissociation as a genuine in-medium effect, as in the case of gluo-dissociation for a heavy quarkonium state~\cite{Brambilla:2008cx,Brambilla:2011sg}. The thermal width from the pure quadrupole contribution reads
\begin{eqnarray}
    \Gamma_{\textrm{bsd}}^{n,\mathcal{Q}} &=&  \int \frac{d^3 \bm{p}}{(2 \pi)^3} \frac{\alpha}{120} |\Delta E_p^n|^5  n_{\textrm{B}}(|\Delta E_p^n|) \left[ |\langle\bm{p} | \bm{r}^2 |n  \rangle|^2 + 2 |\langle\bm{p} | r^i r^j |n  \rangle|^2\right] 
    \nonumber
    \\
    &=&  \frac{\alpha}{240} \int_{|\bm{k}|>|E_n|} \frac{d^3 \bm{k}}{(2 \pi)^3} \,  n_{\textrm{B}} (|\bm{k}|) |\bm{k}|^3 M^{\frac{3}{2}} \sqrt{|\bm{k}|+E_n} 
    \nonumber
    \\
    && \hspace{4 cm} \times \left[ |\langle\bm{p} | \bm{r}^2 |n  \rangle|^2 + 2 |\langle\bm{p} | r^i r^j |n  \rangle|^2\right] \Big \vert_{|\bm{p}|=\sqrt{M(|\bm{k}|+E_n)}} \, , \nonumber \\
    \label{thermal_width_Q_2}
\end{eqnarray}
where $\Delta E_p^n=E_n-E_p=-\Delta E_n^p <0$, and in the last equation we have used the energy conservation $|\bm{k}|+E_n=p^2/M$ to trade the relative momentum of the pair for the momentum of the incoming scalar particle. We notice that the matrix elements of pNRY have to be evaluated accordingly for $|\bm{p}|=\sqrt{M(|\bm{k}|+E_n)}$ fixed by the momentum conservation. Furthermore, one can see that the scalar needs to have a threshold momentum to trigger the \emph{thermal} break-up of the bound state.
 
 Our result in \Eq\eqref{thermal_width_Q_2} resembles expressions obtained for the gluo-dissociation of a color-singlet quark-antiquark in the same hierarchy of scales~\cite{Brambilla:2011sg}, and for the hydrogen atom in QED~\cite{Escobedo:2008sy}. In the form of the second expression, the thermal width \eqref{thermal_width_Q_2} can be interpreted as a convolution of the scalar thermal distribution and an \emph{in-vacuum} function of the scalar momentum. The latter can be taken to be the in-vacuum scalar-dissociation cross section, and we identify it with $\sigma^n_{\textrm{bsd}}(|\bm{k}|)$ as appearing in the factorized form\footnote{To be more precise, the identification would work with the inclusion of a relative velocity, say $v_{\hbox{\scriptsize rel}}'$. However, this is not the relative velocity of the particle-antiparticle pair $v_{\hbox{\scriptsize rel}}$, but rather the relative velocity between a darkonium bound state and a highly relativistic scalar from the thermal bath. Therefore, $v_{\hbox{\scriptsize rel}}'$ can be safely set to unity.} 
 \begin{eqnarray}
     \Gamma_{n} = \int_{|\bm{k}|\geq |E_{n}|} \frac{d^{3}\bm{k}}{(2\pi)^{3}}~n_{B}(|\bm{k}|)  \, \sigma^n_{\textrm{bsd}} (|\bm{k}|) \, .
            \label{dis_width_conv}
     \end{eqnarray}
	 Let us again emphasize that the dissociation rate is obtained from pNRY$_{\gamma_5}$ at finite temperature, by calculating the thermal self-energy of the bound-state. The thermal average of the incoming scalar particle naturally arises from a quantum field theoretical derivation, rather than being inferred from a Boltzmann-like treatment. Indeed, we have never introduced rate equations to formulate thermal rates in the first place. In the context of DM, the dissociation rate from a vector boson (a dark photon for a U(1) Abelian DM model) has been given in \myRef\cite{vonHarling:2014kha}, where the convolution of the in-vacuum dissociation cross section with the thermal distribution of the vector boson has been introduced by relying on a network of Boltzmann-like equations for the heavy pairs. 
  
   In order to provide the full thermal width at the leading order, and the corresponding bound-state dissociation cross section, all the diagrams in figure~\ref{fig:pNRY_width_sec} have to be considered. The calculation is very similar to the case just shown, and from the total thermal width, $\Gamma_{\textrm{bsd}}=\Gamma^{\mathcal{Q}}_{\textrm{bsd}}+\Gamma^{\mathcal{\nabla}}_{\textrm{bsd}}+\Gamma^{\textrm{mix}}_{\textrm{bsd}}$, we write the total
 dissociation cross section as follows 
 \begin{eqnarray}
     &&\sigma^n_{\textrm{bsd}}(|\bm{k}|) = \alpha M^{\frac{3}{2}} \sqrt{|\bm{k}|+E_n}  \left\lbrace \frac{|\bm{k}|^3}{240} \left[ |\langle\bm{p} | \bm{r}^2 |n  \rangle|^2 + 2 |\langle\bm{p} | r^i r^j |n  \rangle|^2\right] \right. 
\nonumber
     \\
     && \hspace{1.5 cm}\left.  +\frac{1}{|\bm{k}|} \Big \vert \Big \langle \bm{p} \Big \vert \frac{\nabla^2_{\bm{r}}}{M^2} \Big \vert n  \Big \rangle \Big \vert^{2}  -\frac{|\bm{k}|}{6} \textrm{Re} \left[ \Big \langle \bm{p} \Big \vert \frac{\nabla^2_{\bm{r}}}{M^2} \Big \vert n  \Big \rangle  \langle n | \bm{r}^2 | p \rangle \right]  \right\rbrace  \Big \vert_{|\bm{p}|=\sqrt{M(|\bm{k}|+E_n)}} \, .
     \label{dis_cross}
 \end{eqnarray}
  \begin{figure}[t!]
         \centering
         \includegraphics[width=0.45\textwidth]{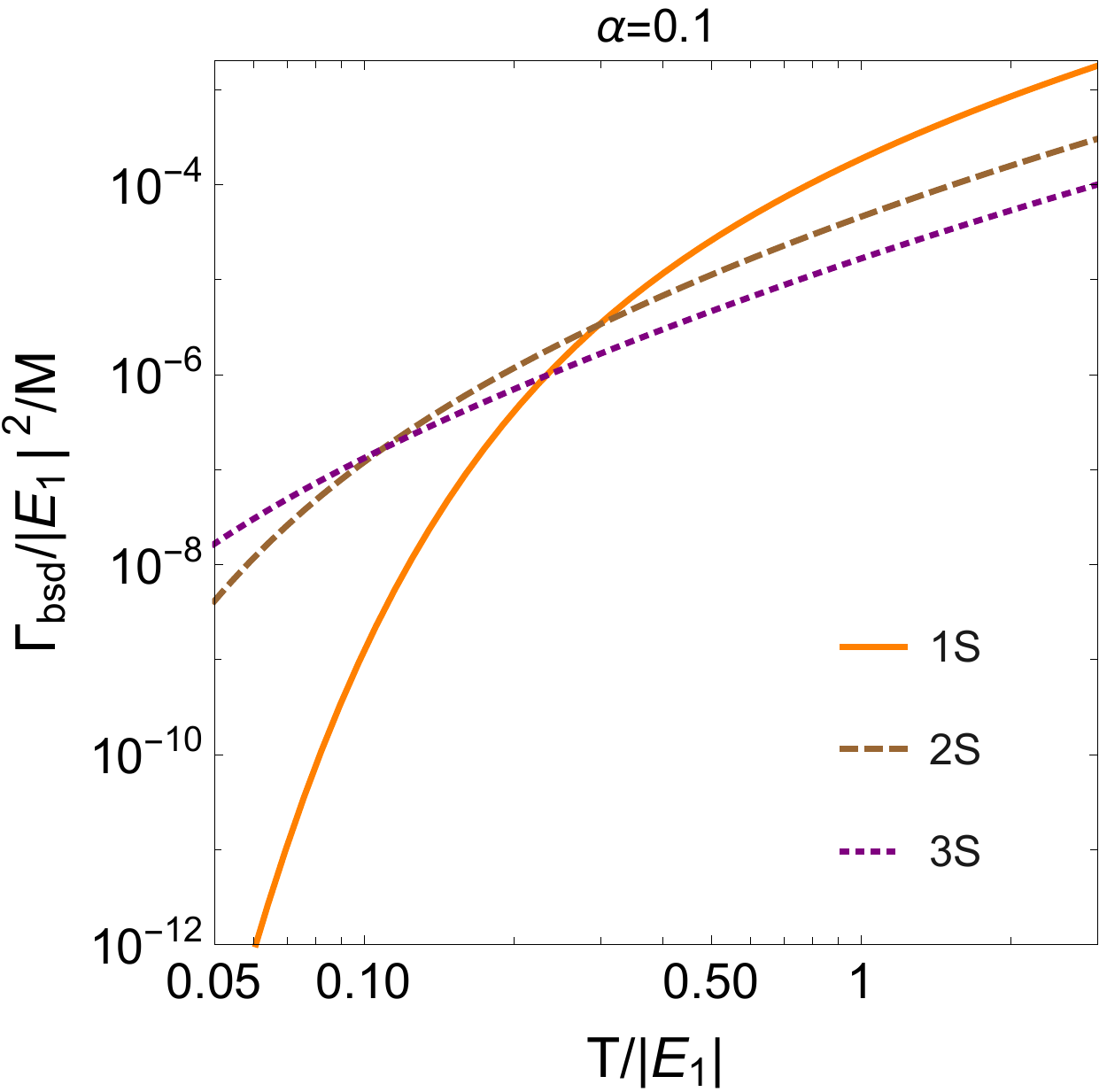}
         \hspace{0.5 cm}
          \includegraphics[width=0.45\textwidth]{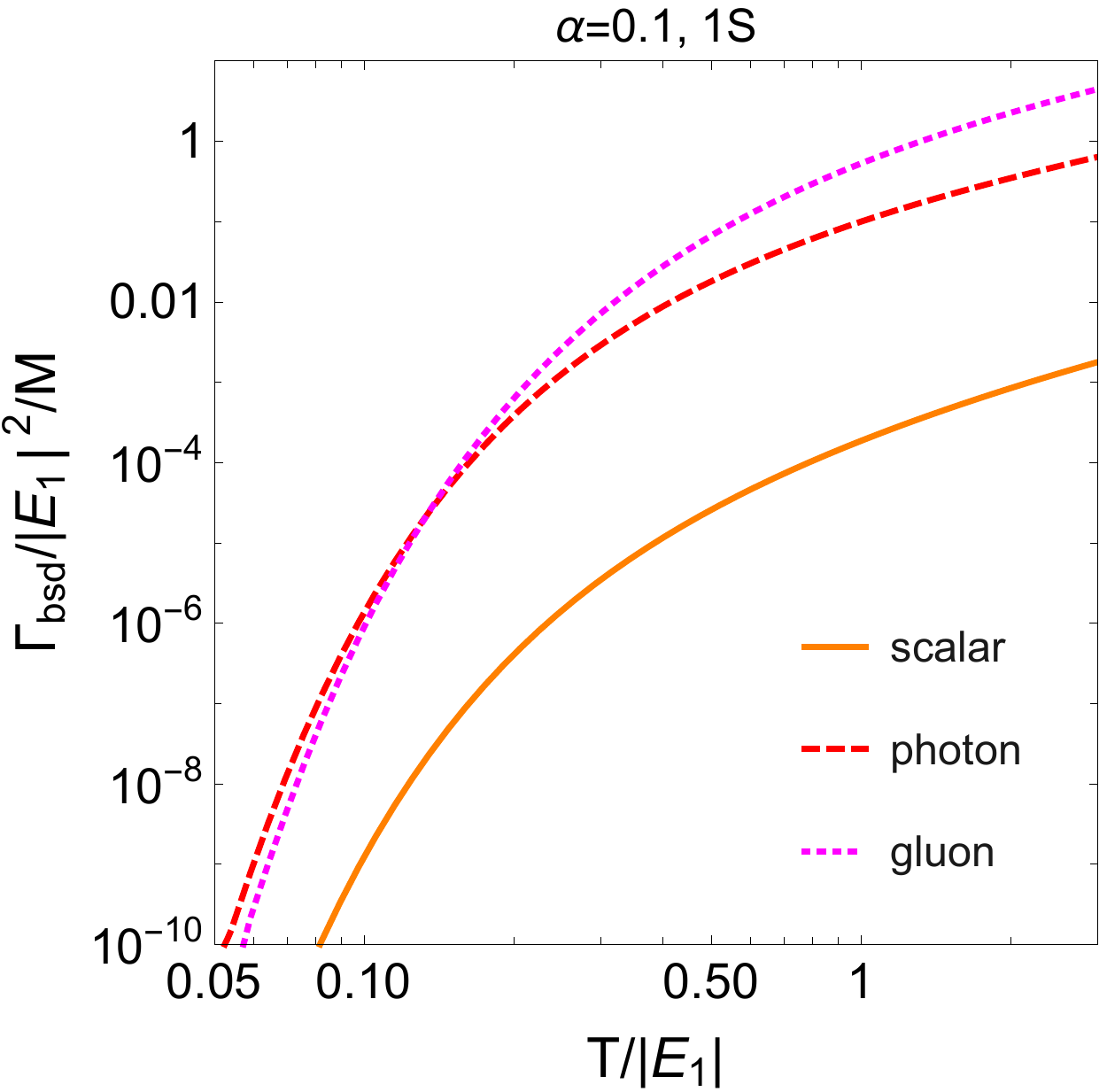}
         \caption{(Left) Thermal width of the bound states, normalized by $|E_1|^2/M$, for the $\ket{1S}$, $\ket{2S}$ and $\ket{3S}$ states. We fixed $\alpha=0.1$. (Right) Comparison among the thermal widths induced by a scalar, photon and gluon (with $N_c=3$ in \Eq\eqref{disso_cross_gluon}) for the ground state.}
         \label{fig:gamma_diss}
     \end{figure}
The result is organized in a transparent way in terms of pNRY matrix elements, that can be evaluated in a quantum mechanical calculation for the given bound state. As a reference, in the following we list the result for the ground state
 \begin{equation}
     \sigma^{1S}_{\textrm{bsd}}(|\bm{k}|)  =
        \frac{ \alpha^3 \, 2^{7} \pi^2}{15} \frac{|E_1|^2}{M|\bm{k}|^3} \left( 7-4 \frac{|E_1|}{|\bm{k}|}\right) \frac{e^{-\frac{4}{w_1(|\bm{k}|)} \arctan (w_1(|\bm{k}|))}}{1-e^{-\frac{2 \pi}{ w_1(|\bm{k}|) }}}\, ,
 \end{equation}
 where $w_1(|\bm{k}|) \equiv \sqrt{|\bm{k}|/|E_1|-1}$.
 Explicit expression for the $2S$, $3S$, $2P$ and $3P$ states can be found in appendix~\ref{app:diss_excited}. In figure~\ref{fig:gamma_diss} we show the dissociation rate for the first $nS$ states. One may notice how for $T \siml 0.3 |E_1|$, the bound-state dissociation is larger for the shallower states. 
 
 By adopting the results available in the literature, we can write the bound-state dissociation cross section induced by a dark photon~\cite{vonHarling:2014kha} for an Abelian gauge group. It reads
 \begin{equation}
        \sigma^{1 \textrm{S}}_{\hbox{\scriptsize bsd}}(|\bm{k}|) \Big \vert_{\textrm{photon}}= \ \frac{\alpha \,  2^{10} \pi^2}{3} \frac{|E_1|^3}{M  |\bm{k}|^4}  \frac{e^{-\frac{4}{w_1(|\bm{k}|)}\arctan(w_1(|\bm{k}|))} }{1-e^{-\frac{2\pi}{w_1(|\bm{k}|)}}} \, .
        \label{disso_cross_photon}
     \end{equation}
     For the sake of comparison, we also add the result for a heavy quarkonium~\cite{Brambilla:2011sg}, that illustrates the case of a non-Abelian vector boson, where the scattering state is in an octet configuration. Then, the unbounded pair in the final state experiences a repulsive interaction
  \begin{equation}
        \sigma^{1 \textrm{S}}_{\hbox{\scriptsize bsd}}(|\bm{k}|) \Big \vert_{\textrm{gluon}}= \ \frac{\alpha C_F \,  2^{10} \pi^2}{3} \rho (\rho+2)^2 \frac{|E_1|^3}{M  |\bm{k}|^4} \left( 1+ \frac{|E_1|}{|\bm{k}|} (\rho^2-1)  \right) \frac{e^{\frac{4 \rho}{w_1(|\bm{k}|)}\arctan(w_1(|\bm{k}|))} }{e^{\frac{2\pi \rho}{w_1(|\bm{k}|)}}-1} \, ,
        \label{disso_cross_gluon}
     \end{equation}
     where $\rho \equiv 1/(N_c^2-1)$, with $N_c$ being the number of colors.
     By comparing the powers of the corresponding fine structure constants, one may see how the bound-state dissociation induced by the thermal scalar is $\alpha^2$-suppressed with respect to the photo- and gluo-dissociation in \Eqs \eqref{disso_cross_photon} and \eqref{disso_cross_gluon} respectively. This can be traced back to the different vertices inducing the ultrasoft or thermal transitions (quadrupole and derivative for the scalar case and dipole interactions for the vector mediators). Upon plugging the dissociation cross sections into \Eq\eqref{dis_width_conv}, the scalar, photo- and gluo-dissociation rates are given in figure~\ref{fig:gamma_diss} (right panel) for the ground state.

\section{Pair annihilations in \texorpdfstring{pNRY$_{\gamma_5}$}{pNRY5}}
 \label{sec:pNRYann}
 As a preparation for the next section where we address the extraction of the DM energy density, let us first discuss the annihilations of heavy pairs. This process is responsible for the depletion of the DM particles into lighter degrees of freedom. Annihilations can equally occur for a particle-antiparticle pair in a scattering state or in a bound state. At variance with the annihilation cross section presented in \Eq\eqref{NR_hard_cross_section_S} for the scattering states, where the heavy fermions were taken as free particles, here we shall include the effect of soft momentum exchange as mediated by the scalar particle $\phi$. In order to derive pNRY$_{\gamma_5}$, we integrate out the momentum scale associated to the soft scalar exchange, thus obtaining the leading order Coulomb potential. Accordingly, the bilocal field $\varphi$ in pNRY$_{\gamma_5}$ satisfies the Schrödinger equation with the potential induced by the scalar exchange. In the evaluation of the annihilation diagrams
 we will also encounter the wave-function of the interacting pair that will lead us to quantum
 mechanical matrix elements entering our quantum field theoretical predictions.
 
 In NRY$_{\gamma_5}$ heavy pair annihilations are accounted for by local four-fermion operators. 
 The same process can be also described in pNRY$_{\gamma_5}$, where the four-fermion operators of NRY$_{\gamma_5}$ generate local terms in the imaginary part of potential $V(\bm{p},\bm{r},\bm{\sigma}_1,\bm{\sigma}_2)$ in \Eq\eqref{pNREFT_sca_0}. As for the NRY$_{\gamma_5}$, the relevant ingredient is the imaginary part of the matching coefficients as given in \Eqs\eqref{match_sca_dim_6} and \eqref{match_sca_dim_8_a}--\eqref{match_sca_dim_8}. As mentioned earlier, the presence of the pseudo-scalar coupling allows for velocity-independent pair annihilations (\cf \Eq\eqref{NR_hard_cross_section_S}) which are typically more relevant for the freeze-out dynamics that fixes the DM energy density. Nevertheless, a non-trivial dependence on the relative magnitude between $g$ and $g_5$ can make velocity-dependent annihilations equally relevant, which is why in the following we explicitly include the corresponding operators. We write the annihilation term from the Lagrangian density of pNRY$_{\gamma_5}$ as follows~\cite{Brambilla:2002nu,Brambilla:2004jw} 
 \begin{eqnarray}
\mathcal{L}^{\textrm{ann}}_{\hbox{\tiny pNRY}_{\gamma_5}}&=& \frac{i}{M^2} \, \int d^3 \bm{r} \varphi^\dagger (\bm{r}) \delta^3(\bm{r}) \left[ 2 {\rm{Im}}[f(^1S_0)] - \bm{S}^2 \left( {\rm{Im}}[f(^1S_0)]-  {\rm{Im}}[f(^3S_1)] \right) \right] \varphi (\bm{r}) 
\nonumber
\\
&&\frac{i}{M^4} \, \int d^3 \bm{r} \varphi^\dagger (\bm{r}) \mathcal{T}_{SJ}^{ij} \nabla_{\bm{r}}^i \delta^3(\bm{r})  \nabla_{\bm{r}}^j \,  {\rm{Im}}  [f(^{2 S+1}P_{J})] \varphi \, (\bm{r}) 
\nonumber
\\
&&\frac{i}{2M^4} \, \int d^3 \bm{r} \varphi^\dagger (\bm{r}) \,  \Omega_{SJ}^{ij} \left\lbrace \delta^3(\bm{r}),\nabla_{\bm{r}}^i \nabla_{\bm{r}}^j  \right\rbrace  {\rm{Im}} [g(^{2 S+1}S_{J})] \varphi \, (\bm{r}) \, ,
\label{pNRY_ann}
 \end{eqnarray}
 where $\bm{S}$ is the total spin of the pair ($\bm{S}^2=0$ for spin singlets and $\bm{S}^2=2$ for spin triplets), while $\mathcal{T}_{SJ}^{ij}$ and $\Omega_{SJ}^{ij}$ are spin projector operators (\cf \eg~\cite{Brambilla:2002nu,Brambilla:2004jw}). We did not write the $\bm{R}$ and $t$ dependence in the argument of the field $\varphi$ to avoid cluttering the notation. Some comments are in order. First, the Lagrangian term given in \Eq\eqref{pNRY_ann} is suited to provide the annihilation of the heavy fermion-antifermion pair both in a scattering or a bound state. The different normalization of the states, as outlined in section~\ref{sec:bsf} and section~\ref{sec:bsd}, will automatically determine the corresponding observable: on the one hand a cross section for the annihilation of scattering states, on the other hand  a bound-state decay width. Second, the operators are contact terms, as inherited from NRY$_{\gamma_5}$, and the wave functions and their derivatives contribute only at the origin ($\bm{r}=\bm{0}$), as originally showed for NRQCD~\cite{Braaten:1996ix}. Rigorous formulations for quarkonium wave functions have been developed in the pNRQCD formalism~\cite{Pineda:1997bj,Brambilla:1999xf,Brambilla:2004jw}.
 
 \begin{figure}[t!]
     \centering
     \includegraphics[width=0.65\textwidth]{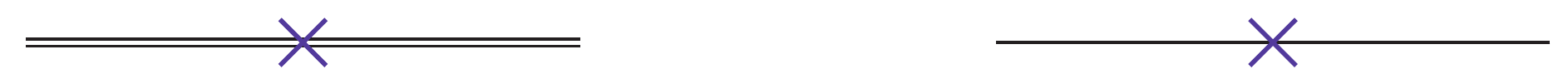}
     \caption{Annihilation diagrams in pNRY$_{\gamma_5}$ for a scattering state (left) and a bound state (right).}
     \label{fig:tree_level_ann}
 \end{figure}
 
 In pNRY$_{\gamma_5}$ we can express the decay width from the self-energy of the field $\varphi$ in the same fashion as we did it for the bound-state dissociation/formation at one loop. This amounts to a simple vertex insertion at tree-level as shown in figure~\ref{fig:tree_level_ann}. Then the decay width for $nS$ bound states can be written as
 \begin{equation}
     \Gamma_{\textrm{ann}}^{nS} = \bra{nS} 2 \textrm{Im} (-\Sigma_{\textrm{ann}}) \ket{nS} \,,
 \end{equation}
where the subscript ``ann'' stands for annihilation. 

At lowest order in the couplings, as one may see from \Eq\eqref{match_sca_dim_6}, only bound-states in a spin-singlet configuration can annihilate into a pair of scalars. Moreover, the operator in the second line of \Eq\eqref{pNRY_ann} does not contribute because the derivative of $nS$ wave functions vanishes at the origin.
 The corresponding decay width reads
  \begin{eqnarray}
  \Gamma_{\textrm{ann}}^{nS} &=& \frac{|R^{(0)}_{nS}(0)|^2}{\pi M^2} \left\lbrace   {\rm{Im}}[f(^1S_0)] + \frac{E^{(0)}_{n}}{M}   {\rm{Im}}[g(^1S_0)]  \right\rbrace  = \frac{M \, \alpha^4 \alpha_5}{n^3} \left( 1+ \frac{\alpha^2}{3 n^2}\right) \, ,
   \label{decay_nS_general}
 \end{eqnarray}
 where in the last step we used $|R^{(0)}_{nS}(0)|^2=4/(n^3 a_0^3)$ from the wave function given in appendix~\ref{sec:appendix-me}, the Bohr radius and the binding energy at leading order from \Eq\eqref{Coulomb_en_levels}. 
 Let us also comment on the second term in the curly brackets in \Eq\eqref{decay_nS_general}. This term is generated by the operator in the third line of \Eq\eqref{pNRY_ann}, and corresponds to the wave function combination $\textrm{Re}(R^\ast_{n S} \nabla^2_{\bm{r}}  R_{n S})(\bm{r})$, that diverges at $\bm{r}=\bm{0}$. As noted in the context of NRQCD, and subsequently reinterpreted in pNRQCD, some perturbative matrix elements are indeed UV divergent and require regularization and renormalization. Here, we employ the relation
$\textrm{Re}(R^\ast_{n S} \nabla^2_{\bm{r}}  R_{n S})(0) = - M E_{nl} |R_{n S}(0)|^2$
from~\cite{Brambilla:2002nu}, which holds in dimensional regularization, and leads to the expression in \Eq\eqref{decay_nS_general}. Alternative possibilities to regularize this divergence employ a hard cut-off~\cite{Braaten:1996ix}, or well dimensional regularization in conjunction with the $\overline{\textrm{MS}}$ scheme~\cite{Chung:2020zqc}.

Let us remark that, in principle, these annihilation rates can be also obtained directly in NRY$_{\gamma_5}$
by relating the bound-state-to-vacuum matrix elements to the wave functions at the origin. The corresponding discussion in the context of NRQCD and heavy quarkonia can be found \eg in section III C of~\cite{Bodwin:1994jh}. In this respect the situation in NRY$_{\gamma_5}$ is even simpler as compared to NRQCD since our matrix elements are perturbative (at least in the Coulomb-limit) and do not contain covariant derivatives. The decays of heavy fermions into two scalars in NRY$_{\gamma_5}$ correspond to the electromagnetic decays of $\eta_{q}$ or $\chi_{qJ}$ with $q=c,b$ into 2 photons in NRQCD.
 
 For the ground-state and the first two excited $2S$ and $3S$ states, one finds from \Eq\eqref{decay_nS_general}, and the matching coefficients in \Eqs \eqref{match_sca_dim_6} and \eqref{match_sca_dim_8_a}--\eqref{match_sca_dim_8}
 \begin{equation}
      \Gamma_{\textrm{ann}}^{1S} = \alpha_5 \alpha^4  M \left( 1 + \frac{\alpha^2}{3} \right) \, , \quad  \Gamma_{\textrm{ann}}^{2S} = \frac{\alpha_5 \alpha^4 M}{8} \left( 1 + \frac{\alpha^2}{12} \right)  \, , \quad  \Gamma_{\textrm{ann}}^{3S} = \frac{\alpha_5 \alpha^4 M}{27} \left( 1 + \frac{\alpha^2}{27} \right)\, .
      \label{decay_width_nS}
 \end{equation}
 The same derivation holds for $nP_J$ states, where the decay width reads
 \begin{eqnarray}
  \Gamma_{\textrm{ann}}^{nP_J} = \frac{3 |R'^{(0)}_{nP}(0)|^2}{\pi M^4}  {\rm{Im}}[f(^3 P_J)] \, , \quad \textrm{for} \; J=0,2 \, .
 \end{eqnarray}
 It is easy to find the result for the two combinations of the total angular momentum $J=0,2$ by using the matching coefficients in \Eq\eqref{match_sca_dim_8_a}-\eqref{match_sca_dim_8} and $|R'^{(0)}_{nP}(0)|^2=4(n^2-1)/(9 n^5 a_0^5)$.
 
 In order to compute the annihilation cross section, we instead evaluate the annihilation vertex on a scatterings state, where we include the spin average factor $1/4$ as it was done in \Eq\eqref{NR_hard_cross_section_S}
 \begin{align}
     \sigma_{\textrm{ann}} v_{\textrm{rel}} & = \bra{\bm{p}} 2 \textrm{Im} (-\Sigma_{\textrm{ann}}) \ket{\bm{p}} =
     \frac{{\rm{Im}}[f(^1S_0)]}{M^2} |\mathcal{R}_{0}(0)|^2 
     \nonumber
     \\
     &+  \frac{{\rm{Im}}[f(^3P_0)]+5{\rm{Im}}[f(^3P_2)]}{3 M^4}   \big \vert \mathcal{R}'^{(0)}_{1}(0) \big \vert^2  + \frac{{\rm{Im}}[g(^1S_0)]}{M^4} \textrm{Re}(\mathcal{R}^*_{0} \nabla^2_{{\bm{r}}} \mathcal{R}_{0} )\big \vert_{r=0} 
     \nonumber
     \\
     &= \frac{2 \pi \alpha \alpha_5}{M^2} \left( 1 - \frac{v^2_{\textrm{rel}}}{3}\right) S(\zeta)  + \frac{(9 \alpha^2 - 2 \alpha \alpha_5 + \alpha_5^2)v_{\textrm{rel}}^2}{24 M^2} S_p(\zeta)    \, ,
     \label{ann_cross_Sommerfeld}
 \end{align}
 where in the second line used the expressions for the radial wave function at the origin for a DM pair in a scattering state (\cf \Eq\eqref{scattering_wave_function}), while the $S$- and $P$-wave Sommerfeld factors are defined as follows
 \begin{eqnarray}
 \big \vert \mathcal{R}^{(0)}_{0}(0) \big \vert^2=\frac{2 \pi \zeta}{1-e^{-2 \pi \zeta}} \equiv S(\zeta) \, , \quad  \big \vert \mathcal{R}'^{(0)}_{1}(0) \big \vert^2 =p^2 S(\zeta) (1+\zeta^2) \equiv p^2 S_p(\zeta) \, ,
 \end{eqnarray}
with $p=M v_{\textrm{rel}}/2$. We remark that also the quantity $\textrm{Re}(\mathcal{R}^*_{0} \nabla^2_{\bm{r}} \mathcal{R}_{0} )\big \vert_{r=0}$ appearing in \Eq\eqref{ann_cross_Sommerfeld} is divergent. We employ the analog expression already exploited for the bound states for the positive part of the spectrum, namely the scattering states. Hence, we trade the divergent combination for a finite expression using $\textrm{Re}(\mathcal{R}^*_{0} \nabla^2_{{\bm{r}}} \mathcal{R}_{0}  )(\bm{0}) =  p^2 |\mathcal{R}_{0}(\bm{0})|^2$.

We would like to highlight that in pNRY$_{\gamma_{5}}$, the annihilation cross sections automatically include the Sommerfeld enhancement originating from the attractive Coulomb potential, namely $S(\zeta)$ and $S_p(\zeta)$ that agree with the literature (\cf \eg\cite{Hisano:2004ds,Cassel:2009wt,Iengo:2009ni}). In other words, when computing the annihilation cross section for scattering states, the resummation of multiple soft-scalar exchanges (ladder diagrams) is already taken care of, since pNRY is a quantum field theory of \emph{interacting pairs}. In the limit $\alpha_5 \to 0$ our \Eq\eqref{ann_cross_Sommerfeld} reproduces the result from \myRef\cite{Baldes:2017gzw} for the same model.

\section{Phenomenological discussion on the energy density}
 \label{sec:pheno}
 The depletion of DM through bound states in the early universe depends on many aspects: which darkonium states are formed, how fast they decay and how efficiently they are dissociated by the interactions with the medium. 
 The thermal rates we have computed in section~\ref{sec:bsf} and section~\ref{sec:bsd} can serve as ingredients when plugged into some rate equations that govern the time evolution of bound and scattering states. Our derivation of thermal cross sections and dissociation rates is based on a quantum field theoretical treatment at finite temperature, meaning that the evolution equations should be worked out in the same fashion. A promising approach that brings together pNREFTs and the evolution equations of heavy pairs in a thermal environment has been recently put forward in the case of heavy quarkonium~\cite{Brambilla:2016wgg,Brambilla:2017zei,Yao:2018nmy}. However, a derivation of quantum evolution equations for DM pairs in a thermal environment is beyond the scope of the present work.\footnote{For DM with an Abelian vector mediator the corresponding evolution equations from pNRQED in an open-quantum-system approach are addressed in a work in preparation \cite{B_and_NAG_2}.}
 
Nevertheless, there already exists a treatment of bound states in terms of semi-classical Boltzmann equations, which allows estimating the bound-state effects on the DM energy density~\cite{vonHarling:2014kha,Ellis:2015vaa}.\footnote{Such treatment has been very recently revisited in \myRefs\cite{Garny:2021qsr,Binder:2021vfo}.} In the most general case, the situation is rather intricate since there is an equation for the single DM particle number density, denoted by $n_{ X}$, and an equation for the number density of  each bound state, denoted by $n_{\varphi_b}$. All possible reactions have to be taken into account. Yet, under the assumption that the bound states are kept close to the chemical equilibrium through their direct and inverse decays, that is ensured by having $\Gamma_{\textrm{ann}} \gg H$, where $H$ is the Hubble rate, one can obtain algebraic equations for each $n_{\varphi}$ in terms of its equilibrium counterpart $n^{ \textrm{eq}}_{\varphi_b}$ and $n_X$, and reuse these results in the Boltzmann equation for $n_X$. Such an approach was  originally suggested  in~\cite{Ellis:2015vaa}, and it is often adopted in the literature on the subject. Here, the annihilations of DM particles are treated in an effective way with a single Boltzmann equation for the DM number density $n_{\hbox{\tiny X}}$ which reads
 \begin{equation}
     \frac{d n_{X}}{d t} + 3H n_{X} = - \langle \sigma_{\textrm{eff}} \, v_{\textrm{rel}} \rangle (n^2_{X}-n^2_{X,\textrm{eq}}) \, ,
     \label{Boltzmann_eq_eff}
 \end{equation}
 where the Hubble rate can be expressed in terms of the energy density $H=\sqrt{8 \pi e /3}/M_{\textrm{Pl}}$, 
 where $M_{\textrm{Pl}}$ is the Planck mass with $M_{\textrm{Pl}} = 1.22 \times 10^{19}\textrm{ GeV}$, while $e=\pi^2 T^4 g_{\textrm{eff}}/30$ and $g_{\textrm{eff}}$ denotes the effective number of  relativistic degrees of freedom.
The effective thermally averaged cross section, when neglecting bound-to-bound transitions, is
 \begin{equation}
    \langle  \sigma_{\textrm{eff}} \, v_{\textrm{rel}} \rangle  =  \langle \sigma_{\textrm{ann}} \, v_{\textrm{rel}} \rangle + \sum_{n} \langle   \sigma^n_{\textrm{bsf}} \, v_{\textrm{rel}} \rangle \, \frac{\Gamma_{\textrm{ann}}^n}{\Gamma_{\textrm{ann}}^n+\Gamma_{\textrm{bsd}}^n} \, , 
    \label{Cross_section_eff}
 \end{equation}
 where the sum runs over bound states. 
 
 The quantities involved are the thermally averaged  annihilation cross section for the pair in a scattering state $\langle \sigma_{\textrm{ann}} \, v_{\textrm{rel}} \rangle$,  the thermally averaged bound-state formation cross section $\langle   \sigma^n_{\textrm{bsf}} \, v_{\textrm{rel}} \rangle$, the state bound-state decay width $\Gamma_{\textrm{ann}}^n$  and the bound-state thermal width $\Gamma_{\textrm{bsd}}^n$, the latter accounting for the dissociation process. As for the cross sections, we are going to use the standard definition of the thermal average (\cf \eg\cite{Feng:2010zp,vonHarling:2014kha}). Then, the combination of the decay width and dissociation width, that we label with $\mathcal{I}_n \equiv \Gamma_{\textrm{ann}}^n$ /($\Gamma_{\textrm{ann}}^n + \Gamma_{\textrm{bsd}}^n)$, determines to what extent DM annihilations via bound states are efficient. One typically has to wait until the temperature, that sets the scale for the energy of the light particles that hit the bound states, is of order of the binding energy of the bound states or smaller. A small population of ionized bound states  corresponds to  $\Gamma_{\textrm{bsd}} \ll \Gamma_{\textrm{ann}}$ and therefore to $\mathcal{I}_n \simeq 1$. A more quantitative assessment is model dependent, and one has to look at the rates that appear in \Eq\eqref{Cross_section_eff}. In the following, we shall consider values of the pseudoscalar coupling that satisfy $g_5 < g$ according to our discussion in section~\ref{sec:pNR}. 
  \begin{figure}[t!]
     \centering
     \includegraphics[width=0.45\textwidth]{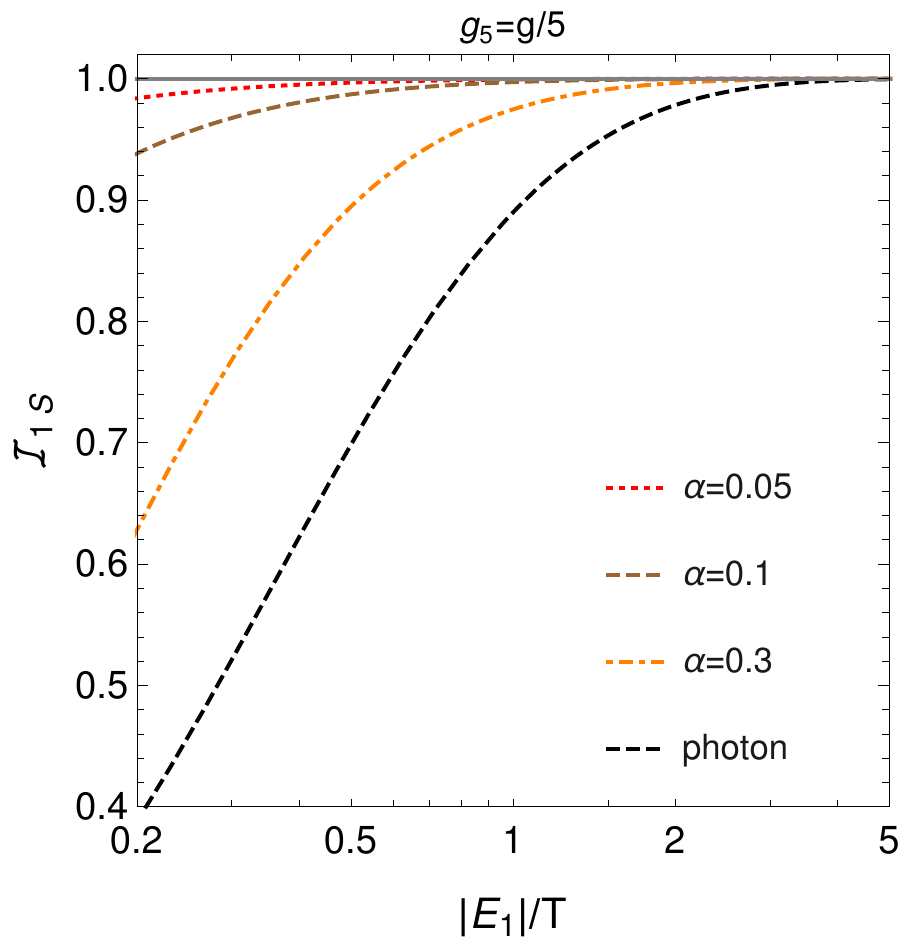}
    \hspace{0.4 cm}
   \includegraphics[width=0.45\textwidth]{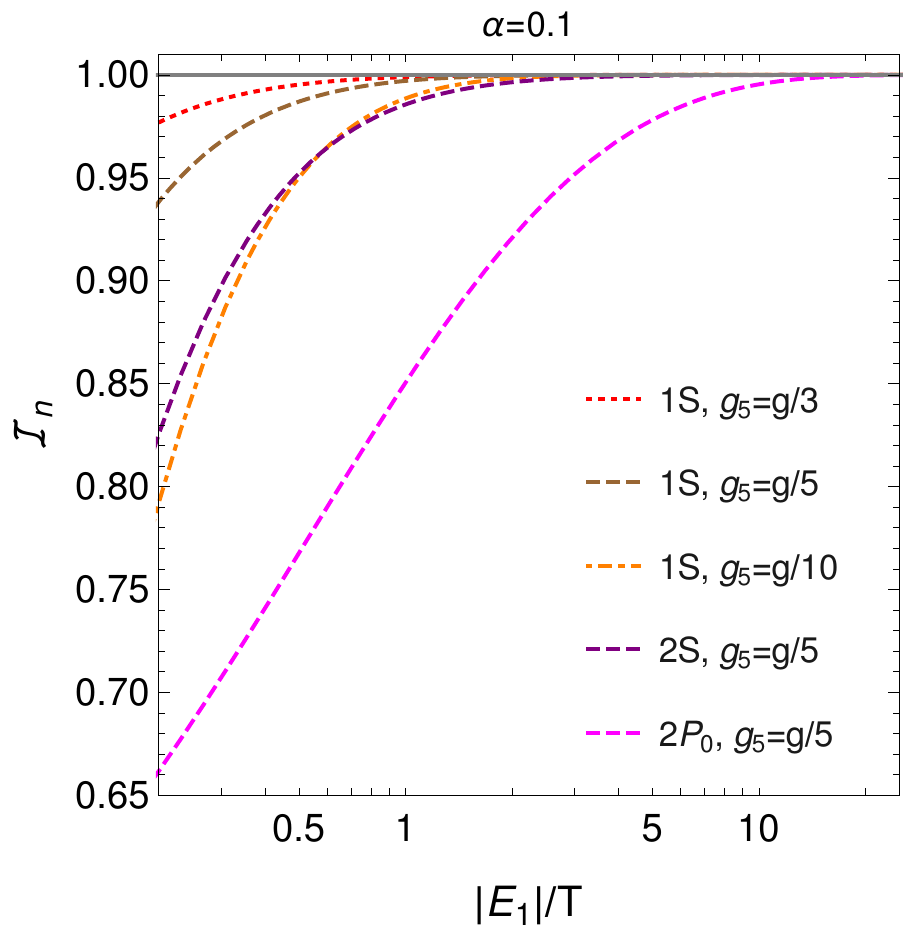}
     \caption{Left: $\mathcal{I}_n$ for the scalar mediator for three values of $\alpha=0.05,0.1,0.3$ displayed with the dotted-red, dashed-brown and dotted-dashed orange curves respectively. The black-dashed line corresponds to the $\mathcal{I}_n$ for the vector mediator model. Right: $\mathcal{I}_n$ for the ground state, for different values of $g/g_5$, and comparison with the $2S$ and $2P_0$ bound states. }
     \label{fig:fig_IR_1S}
 \end{figure}
 
 For the model under consideration benchmark values for $\mathcal{I}_n$ are summarized in figure~\ref{fig:fig_IR_1S} for the ground state.
 Three benchmark values of the coupling constant are chosen as $\alpha=0.05, 0.1, 0.3$ for a fixed combination  $g_5=g/5$. In addition to that, in the right panel, we fix $\alpha=0.1$ and vary the ratio of the coupling $g_5/g$ for the ground state. Let us observe that in the scalar case, there is a visible dependence on $\alpha$. This is different with respect to the case of the vector mediator model in \myRef\cite{vonHarling:2014kha} (black-dashed line in the left panel of figure~\ref{fig:fig_IR_1S}). In this latter case, the combination of the dissociation rate and bound-state decay is such, that the resulting ionization factor is $\alpha$-independent.\footnote{For the bound-state decay width we use $\Gamma_{\textrm{ann}}^{1S}=M \alpha^5/2$, that corresponds to the para-positronium. For this comparison, we take the decay width in \Eq\eqref{decay_width_nS} at leading order, namely at $\mathcal{O}(\alpha_5 \alpha^4)$.} Also, for the model under study,  we obtain larger values of $\mathcal{I}_{1S}$ with respect to the Abelian vector mediator, pointing to an earlier contribution of bound-state effects. Nevertheless, one has to bear in mind that the bound-state formation process mediated by a scalar mediator features an  $\alpha^2$ suppression~\cite{Petraki:2015hla}, and so the overall impact of bound-state effects on the DM energy density is a non-trivial combination of $\mathcal{I}_{n}$ and $\langle \sigma_{\textrm{bsf}} v_{\textrm{rel}} \rangle$ for different models.

 In the right panel of figure~\ref{fig:fig_IR_1S}, one may see how decreasing values of $\alpha_5$, that make the bound-state decay width in \Eq\eqref{decay_width_nS} smaller, but at the same time leaving unaffected the dissociation thermal width, determines a larger ionized population of bound states in the same temperature window. The dashed lines correspond to different states ($1S$, $2S$ and $2P_0$) for the same value of $g_5=g/5$. Excited states remain ionized for longer times. 
 \begin{figure}[t!]
    \centering
    \includegraphics[width=0.45\textwidth]{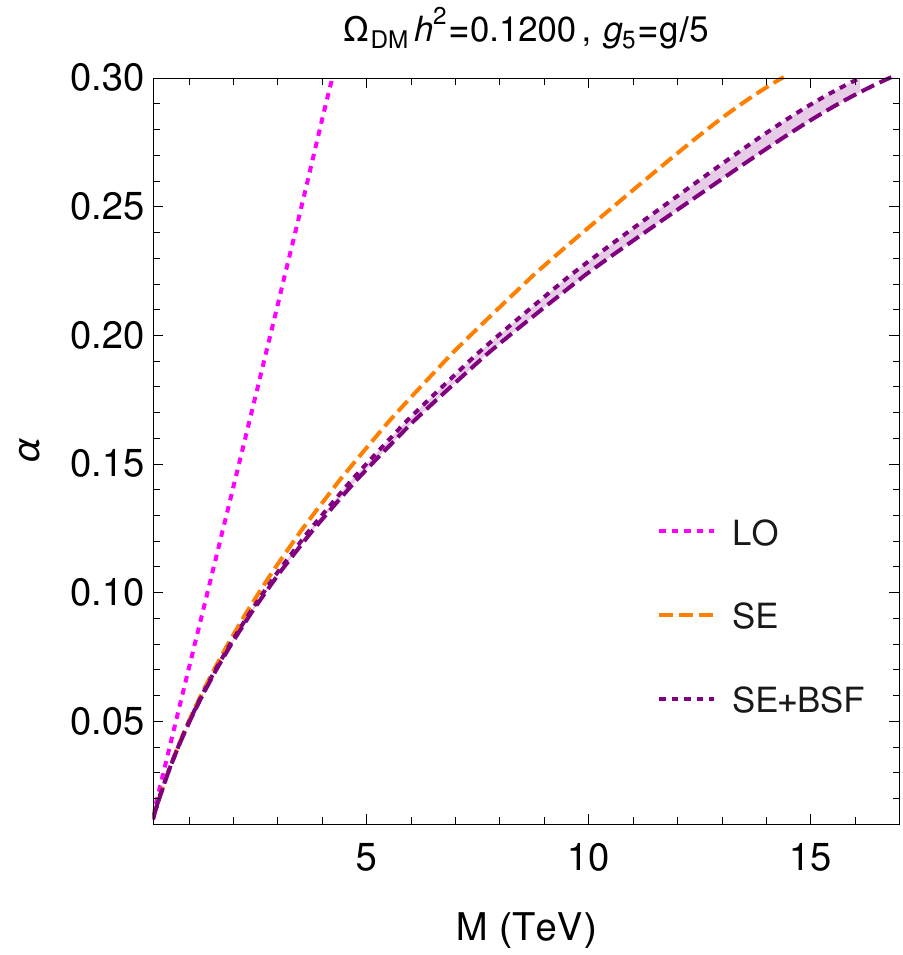}
    \hspace{0.5 cm}
   \includegraphics[width=0.45\textwidth]{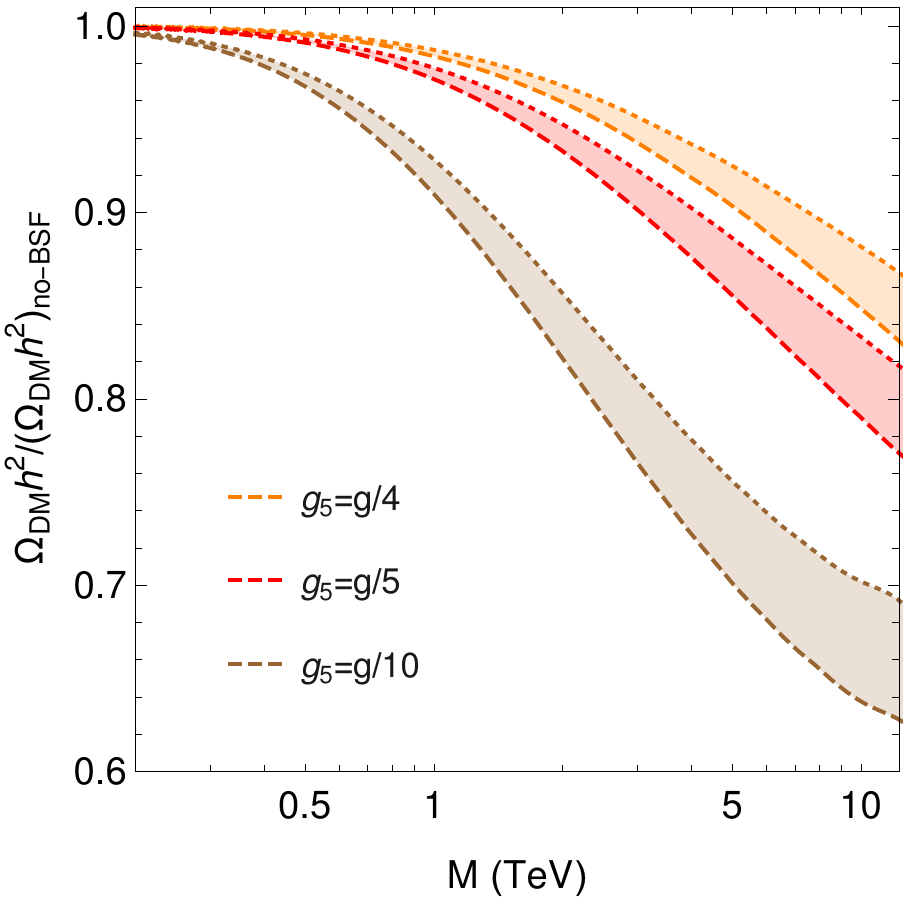}
    \caption{Left: parameter space in the $(M,\alpha)$-plane that reproduces the observed DM energy density for $g_5=g/5$. The purple dotted curve accounts for the ground state only, whereas the purple-dashed curve corresponds to the inclusion of excited states. Right: ratio between the DM energy density as obtained with and without bound-state effects. The parameter $\alpha$ is fixed as a function of the mass $M$ to account for the observed DM energy density when bound-state effects are considered. Dashed (dotted) lines do (not) include excited states. }
    \label{fig:num_Boltzmann}
\end{figure}
 
As a final summarizing result, we give the solution of the effective Boltzmann equation \eqref{Boltzmann_eq_eff}. As usual, we solve the evolution equation for the DM yield $Y_X=n_X/s$, where $s=2 \pi^2 h_{\textrm{eff}}T^3/45$ is the entropy density; we assume that the dark states are at the same temperature as the plasma of SM
particles. The thermalization condition can be easily satisfied via interactions between the scalar mediator and the Higgs boson. The typical size of the couplings that ensure thermalization at high temperatures are typically much smaller than the couplings $g$ and $g_5$ that we consider in this work. Therefore, the corresponding contribution to dark matter annihilations and bound-state decays can be safely neglected. As for $ h_{\textrm{eff}}$ and $ g_{\textrm{eff}}$, we take their temperature-dependent values from \myRef\cite{Laine:2015kra}. In the left plot of figure~\ref{fig:num_Boltzmann}, the curves reproduce the observed DM energy density $\Omega_{\textrm{DM}}h^2 = 0.1200 \pm 0.0012$~\cite{Aghanim:2018eyx} in the parameter space $(M,\alpha)$. Both the Sommerfeld enhancement and bound-state effects are more prominent for large values of $\alpha$ as expected. One may notice the large impact upon including the Sommerfeld enhancement, both for the $S$- and $P$-wave annihilations (\cf \Eq\eqref{ann_cross_Sommerfeld}), with respect to the free annihilation cross section (\cf \Eq\eqref{NR_hard_cross_section_S}). In contrast to that, the contributions of the bound-state formation and decays are moderately relevant, albeit still non-negligible.

In order to single out the bound-state effects, we consider the ratio between the DM energy density as obtained by including or omitting the bound-state formation and decays (namely the second term on the right-hand side of \Eq\eqref{Boltzmann_eq_eff}), in addition to the Sommerfeld enhancement for the pair annihilations. This is shown in the right plot of figure~\ref{fig:num_Boltzmann}. The ratio is smaller than unity because, when neglecting bound-state formation, a smaller population of DM particles is annihilated away, and a larger energy density is found. The formation of bound-states and their decays into light scalars act as an additional channel for depleting DM. 
We notice that a smaller pseudoscalar coupling makes the bound-state formation term more relevant, and this traces back to the different powers of $\alpha$ and $\alpha_5$ entering various cross sections.\footnote{For the case when only the ground state is considered, and looking at the $S$-wave annihilations only, the effective cross section is $\langle \sigma_{\textrm{eff}} v_{\textrm{rel}} \rangle \approx 2 \pi \alpha \alpha_5 \langle S(\zeta) \rangle /M^2 + \pi \alpha^4 \langle S_{\textrm{bsf}}(\zeta) \rangle \mathcal{R}_{1S} /M^2 $. Here, $\langle S_{\textrm{bsf}}(\zeta) \rangle$ can be read off from \Eqs \eqref{bsf_matrix_element_T} and \eqref{bsf_1S}.} For the smaller ratio  between the pseudoscalar and scalar couplings we consider, $g_5=g/10$, and the largest mass compatible with the observed relic density for $\alpha_{\textrm{max}}=0.3$, namely $M=13\,\textrm{TeV}$, we find a 30\%-36\% effect on the predicted energy density if bound-state formation is not included. The dotted lines correspond to the inclusion of the ground state only, whereas upon adding the excited states ($nS$ and $nP$ states with $n=2,3$) we obtain the dashed lines. Bound states with $n>4$ induce a contribution smaller than the uncertainty in the observed DM energy density in our numerical implementation. 

Upon solving the Boltzmann equation \eqref{Boltzmann_eq_eff}, we start the integration at temperatures larger than those strictly admitted by the hierarchy of scales \eqref{hiearchy}, which implies that $T \ll M \alpha$ is not always satisfied. This would amount to deriving the bound-state formation and dissociation rate at such temperatures, in the so-called screening regime, \cf \myRefs\cite{Brambilla:2008cx,Escobedo:2008sy} for QCD and QED, and \myRefs\cite{Biondini:2018pwp,Binder:2018znk} for a DM application.\footnote{Pair annihilations would be affected in terms of the thermal potential at such temperatures, and the corresponding modifications to the wave functions from a plasma-modified Schrödinger equation, \cf \myRefs\cite{Kim:2016kxt,Biondini:2017ufr,Biondini:2018pwp,Binder:2018znk,Kim:2019qix}. From an EFT perspective, one has to integrate out the temperature scale before obtaining a pNREFT, the latter finally attained when removing potential scalars.}  However, due to our assumption for the self-coupling of the scalar, and upon neglecting the explicit interactions in $\mathcal{L}_{\textrm{portal}}$, a thermal mass for the scalar mediator does not appear and, moreover, $2 \to 2$ scattering processes with the medium constituents mediated by a soft scalar are absent at least in the Hard Thermal Loop limit~\cite{Thoma:1994yw}. Despite of the fact that bound-state formation and corresponding decays are more effective at later temperature stages, further investigations on this aspect can be worthwhile, and are left for future research on the subject.

\section{Conclusions}
\label{sec:concl}
In this paper we have taken the first step towards the generalization of a potential non-relativistic effective theory for scalar mediators at finite temperature. In addition to the dynamically generated energy scales typical of bound states, there are also thermodynamic scales at play. Most notably, these are the plasma temperature and thermal masses. Our main goal is to contribute to the recent effort and the development of theoretical tools for estimating the near-threshold and bound-state effects on DM freeze-out calculations. In this scenario DM particles are indeed non-relativistic and slowly moving in the early universe plasma in the relevant temperature window. This means that suitable NREFTs can help in inspecting the corresponding dynamics and calculating thermal cross sections and widths. 

We considered a simplified DM model where the dark particles are Dirac fermions and antifermions interacting via a light scalar mediator. With respect to our previous work, we have included the pseudo-scalar coupling between the mediator and the dark fermions. This very fact has important consequences for the processes that drive the thermal freeze-out of the dark fermions, and ultimately on the corresponding phenomenology. In particular, velocity independent pair annihilations are possible. We recast the relevant processes in terms of low-energy theories, where the degrees of freedom are heavy DM pairs, both in a scattering and in a bound states, as well as ultrasoft and thermal scalars. In our framework, the bound-state formation, bound-state dissociation, annihilations of the unbound pairs and decays of bound states are inferred from (thermal) self-energies of the bilocal fields in pNRY$_{\gamma_5}$. We investigated a specific hierarchy of scales and set of choices for the couplings, that allow us to neglect thermal masses in our present study.

According to our assumed hierarchy of scales, the bound-state formation cross section factorizes into an in-vacuum contribution and a thermal factor. The latter arises from the thermal character of the scalar mediator (\cf \Eq\eqref{bsf_matrix_element_T} and \Eq\eqref{bsf_matrix_element_T0}). The thermal factor implements and resembles the Bose enhancement of the emitted scalar in the formation of a bound-state through radiative emission of the mediator.

A special effort was dedicated to obtain explicit analytic results for the matrix elements of the corresponding pNREFT under the assumption of Coulombic wave functions. We have derived the bound state formation for the ground state, and the excited states in $nS$- and $nP$-wave configurations for the principal quantum numbers $n=2,3$.

The thermal width for the bound state, namely its thermal break-up as induced by a thermal scalar particle, was extracted 
from the thermal self-energy of the corresponding wave function field. It is worth noting that this approach features a manifest factorization between an in-vacuum dissociation cross section and the thermal distribution of the scalar particles in the medium (\cf \Eq\eqref{dis_width_conv}). Most importantly, the in-vacuum dissociation cross section can be written in terms of generic pNRY$_{\gamma_5}$ matrix elements. These quantities represent genuine quantum mechanical expectation values that, depending on the states under consideration and the assumed nature of the potential, can be evaluated by solving the corresponding Schrödinger equation
analytically or numerically. The scalar-induced thermal break up of a bound state is the analog of the photo- and gluo-dissociation, the latter being particularly relevant for heavy quarkonium physics in heavy ion collisions. 

As third and last piece of information, we considered scattering state annihilations and bound state decays, that were written in terms of local operators in pNRY$_{\gamma_5}$. Here the matrix elements arising in the course of the calculation can be naturally understood as expectation values of quantum mechanical operators, where the wave functions follow from solving Schrödinger equation with the corresponding pNRY$_{\gamma_5}$ potential. This is an obvious advantage as compared to NRY$_{\gamma_5}$, where the connection between the given matrix element and the corresponding quantum mechanical expression is not always straightforward. In this respect the situation in pNRY$_{\gamma_5}$ is much more clear. First, in the case of an unbound above-threshold pair, the Sommerfeld enhancement shows up when computing the annihilation cross section. Second, for the negative part of the spectrum, namely bound states, the bound-state decay width is recovered. As an original result of our work, we provided the Sommerfeld enhanced annihilation cross section for the model in \Eq\eqref{lag_mod_0} at  $\mathcal{O}(\alpha^2 v_{\textrm{rel}}^2,\alpha \alpha_5 v_{\textrm{rel}}^2,\alpha_5^2 v_{\textrm{rel}}^2)$. 

We exploit the so-obtained rates in an effective Boltzmann equation, routinely used in phenomenological studies, in order to estimate the effect of the Sommerfeld enhancement, bound-state formation and decays on the DM energy density. We find that a large effect is played by the Sommerfeld enhanced annihilations for the scattering states. In this model, that features a non-vanishing pseudoscalar coupling, $S$-wave velocity-independent annihilations are possible, in addition to the $P$-wave annihilations of the sole scalar interaction case. The former boost DM annihilations in a prominent way, and the DM mass compatible with the observed relic density is pushed to much larger values with respect to those obtained with free annihilations for the same values of the couplings. Moreover, despite the bound-state effects being more moderate, we find them to be non-negligible. Their impact depends on the ratio $g_5/g$ between the couplings in the Lagrangian model \eqref{lag_mod_0}. In particular, decreasing the ratio $g_5/g$ makes bound-state formation more prominent. For $g_5/g=0.25$ ($g_5/g=0.1$) we find a 16\% (35\%) effect on the DM energy density with respect to the case one includes only the Sommerfeld enhancement for the scattering states. The effect of the excited state decays accounts for few per-cents depending again on the particular values of the couplings. A highlight of our study is the observation, that even when the bound-state formation is driven by a scalar mediator, bound-state effects may affect the determination of the DM energy for the simplified model \eqref{lag_mod_0}. The main reason for this is rooted in the presence of a pseudoscalar interaction, and in a non-trivial interplay between bound-state decay widths, bound-state formation and dissociation, and Sommerfeld enhancements. 

We conclude by remarking that, in order to make connections with experimental constraints on the parameter space of the model, one has to specify the portal interactions and inspect the corresponding possible modifications on the various ingredients presented in our study.

\section*{Acknowledgments}
The work of S.\,B. is supported by the Swiss National Science Foundation under the Ambizione grant PZ00P2\_185783. V.\,S. acknowledges the support from the DFG under grant 396021762 -- TRR 257 ``Particle Physics Phenomenology after the Higgs Discovery.'' The authors are grateful to Nora Brambilla and Antonio Vairo for useful discussions on the wave function at the origin in pNRQCD.

\appendix
\numberwithin{equation}{section}

\section{Tree-level Lagrangian for \texorpdfstring{NRY$_{\gamma_5}$}{NRY5}}
\label{sec:appendix-nry}
The bilinear part of the NRY$_{\gamma_5}$ Lagrangian given in \Eqs\eqref{bilinear_psi_5}--\eqref{bilinear_chi_5} can be explicitly obtained from the full theory Lagrangian \Eq\eqref{lag_mod_0} by means of the Foldy-Wouthuysen-Tani (FWT) technique. This way we can directly get the two-fermion operators including their
tree-level matching coefficients. The well known drawbacks of this approach are that the so-obtained operator basis may not be the most useful one and that by construction we miss operators whose Wilson coefficients happen to be loop-induced. Here we would like to refer to appendix A.3 of \myRef\cite{Biondini:2021ccr} for a detailed treatment of the purely scalar Yukawa theory. In the following we will make use of the concepts and the notation introduce there.

The leading order ansatz $\hat{S}$ for the mixed case at hand reads
\begin{equation}
\hat{S}  = - \frac{i}{2 M} \beta \vec{\alpha} \cdot \hat{\bm{p}} + \frac{g_5}{2 M} \gamma_5 \phi,
\label{eq:fwt-ansatz}
\end{equation}
while the Hamiltonian is given by
\begin{equation}
\hat{H} = \vec{\alpha} \cdot \hat{\bm{p}} + M \beta + g \beta \phi + i g_5 \beta \gamma_5 \phi \, .
\end{equation}
It is worth noting that the ansatz given in \Eq\eqref{eq:fwt-ansatz} is identical to $\hat{S}$ required for the pure pseudoscalar Yukawa theory. When doing the decoupling of even and odd operators we need to keep in mind that $\beta \gamma_5$  and $\alpha^{i_1} \ldots \alpha^{i_{2n}} \gamma_5$ with $n \in \mathbb{N}$ are odd, while 
$\alpha^{i_1} \ldots \alpha^{i_{2n+1}} \gamma_5$ is even.

At $\mathcal{O}(1/M)$ the bilinear Lagrangian is just a direct sum of the non-relativistic Lagrangians
for the pure scalar~\cite{Luke:1996hj,Luke:1997ys} and pseudoscalar~\cite{Platzman:1960dqa} Yukawa theories. 
Operators with tree-level matching coefficients induced by the mixing of the two interactions start occurring at $\mathcal{O}(1/M^2)$. For the sake 
of completeness, let us explicitly provide the $\mathcal{O}(1/M^2)$-piece of the bilinear part of the NRY$_{\gamma_5}$ Lagrangian with tree-level  matching coefficients obtained using FWT transformations
\begin{align}
\mathcal{L}^{\textrm{bilinear}}_{\psi, \mathcal{O}(\frac{1}{M^2})} = \psi^\dagger \biggl ( &-\frac{i g \bm{\sigma} \cdot (\bm{\nabla}\phi \times \bm{\nabla})}{4 M^2}-\frac{g  \{ \bm{\nabla}, \{ \bm{\nabla}, \phi \} \} }{8 M^2} -\frac{i g_5 \bm{\sigma} \cdot \{ \bm{\nabla}, (\partial_0\phi) \}}{4 M^2}  \nonumber \\
& +\frac{3 g g_5 \bm{\sigma} \cdot [ \phi^2, \bm{\nabla} ]}{8 M^2}+\frac{g g_5^2 \phi^3}{2 M^2} \biggl )\psi \nonumber \\
\chi^\dagger \biggl ( &\frac{i g \bm{\sigma} \cdot (\bm{\nabla}\phi \times \bm{\nabla})}{4 M^2}+\frac{g  \{ \bm{\nabla}, \{ \bm{\nabla}, \phi \} \} }{8 M^2} +\frac{i g_5 \bm{\sigma} \cdot \{ \bm{\nabla}, (\partial_0\phi) \}}{4 M^2}  \nonumber \\
& +\frac{3 g g_5 \bm{\sigma} \cdot [ \phi^2, \bm{\nabla} ]}{8 M^2}-\frac{g g_5^2 \phi^3}{2 M^2} \biggl )\chi,
\end{align}
where in $(\partial_0\phi)$ the time derivative acts only on the $\phi$ field, while $\bm{\nabla}$ always acts on everything to its right.

\section{Derivation of \texorpdfstring{pNRY$_{\gamma_5}$}{pNRY5} from \texorpdfstring{NRY$_{\gamma_5}$}{NRY5}}
\label{sec:appendix-pnry}
In section~\ref{sec:pNR} it was stated that pNRY$_{\gamma_5}$ can be obtained from
NRY$_{\gamma_5}$ by projecting the Hamiltonian of NRY$_{\gamma_5}$ onto the particle-antiparticle 
sector. Since the technicalities behind this approach are rarely discussed in the
literature, here we would like to provide some more details on the corresponding calculation
for the bilinear part of the NRY$_{\gamma_5}$ Lagrangian.

Switching to $\mathcal{H}_{\gamma_5}$ effectively means multiplying the expressions given in \Eqs\eqref{bilinear_psi_5} and \eqref{bilinear_chi_5} by $-1$ and removing the kinetic terms with time derivatives $\psi^\dagger i \partial_0 \psi$ and $\chi^\dagger i \partial_0 \chi$. Then, we must sandwich
the Hamiltonian between the following ket and bra states
\begin{subequations}
\begin{align}
	\int d^3 \bm{x}_1 d^3 \bm{x}_2 \, \varphi_{ij}(t,\bm{x}_1, \bm{x}_2) \psi^\dagger_i (t,\bm{x}_1) \chi_j (t,\bm{x}_2) \ket{ \phi_{\hbox{\tiny US}}} &\equiv \ket{P},
 \\
	\int d^3 \bm{x}_1 d^3 \bm{x}_2 \bra{ \phi_{\hbox{\tiny US}}}  \chi^\dagger_j (t,\bm{x}_1) \psi_i (t,\bm{x}_2)
		\varphi^\dagger_{ij}(t,\bm{x}_2, \bm{x}_1)  &\equiv \bra{P}.
\end{align} 
	\label{eq:ket1}
\end{subequations}
In order to evaluate the resulting expression we need to move all $\psi$- and 
$\chi^\dagger$- fields to the very right and all $\chi$- and $\psi^\dagger$-fields to the very left, since
\begin{equation}
\psi_i (t, \bm{x}) \ket{ \phi_{\hbox{\tiny US}}} = \chi^\dagger_i (t, \bm{x}) \ket{ \phi_{\hbox{\tiny US}}} =
 \bra{ \phi_{\hbox{\tiny US}}} \psi^\dagger_i (t, \bm{x}) = \bra{ \phi_{\hbox{\tiny US}}} \chi_i (t, \bm{x})  = 
 0.
\end{equation}
To this end we make use of the equal-time anticommutation relations obeyed by the Pauli fields
\begin{equation}
	\{ \psi_i (t, \bm{x}),  \psi_j^\dagger (t, \bm{x}') \} = \{ \chi_i^\dagger (t, \bm{x}) , \chi_j (t, \bm{x}'),  \} = \delta_{ij} \, \delta^{(3)} (\bm{x} - \bm{x}').
\end{equation}
In the case of spatial derivatives acting on one of the fields, it is convenient to let the derivative formally act on the Dirac delta \eg as in
\begin{align}
	\{ \bm{\nabla} \psi_i (t, \bm{x}),  \psi_j^\dagger (t, \bm{x}') \} &= \delta_{ij} \, \bm{\nabla}_{\bm{x}} \delta^{(3)} (\bm{x} - \bm{x}').
\end{align}
When evaluating the integral over $\bm{x}$ we can always integrate by parts to move $\bm{\nabla}_{\bm{x}}$ away from $\delta^{(3)} (\bm{x} - \bm{x}')$.

The bilocal field $\varphi_{ij}$ carries two Pauli indices and behaves as 
\begin{equation}
	\varphi_{ij}(t,\bm{x}_1, \bm{x}_2) \sim \psi_i (t, \bm{x}_1) \chi^\dagger_j (t, \bm{x}_2).
	\label{eq:biloc}
\end{equation}	
Its Hermitian conjugate is given by\footnote{The additional minus sign as compared to the standard relation for the Hermitian conjugate of a matrix $M_{ij} \to M^\dagger_{ji}$ arises from the fact that the elements of 
	$\varphi_{ij}$ are made of anticommuting fields \ie they are Grassmann numbers rather than $c$-numbers.}
\begin{equation}
	\varphi_{ij}(t,\bm{x}_1, \bm{x}_2) \to - \varphi^\dagger_{ji}(t,\bm{x}_1, \bm{x}_2) \sim
	[\psi_i (t, \bm{x}_1) \chi^\dagger_j (t, \bm{x}_2)]^\dagger = -\chi_j (t, \bm{x}_2) \psi^\dagger_i (t, \bm{x}_1),
\end{equation}	
so that
\begin{equation}
	\varphi^\dagger_{ij}(t,\bm{x}_1, \bm{x}_2) \sim \chi_i (t, \bm{x}_2) \psi^\dagger_j (t, \bm{x}_1),
	\label{eq:bilocdagger}
\end{equation}	
and consequently
\begin{subequations}
\begin{align}
	\varphi^\ast_{ij}(t,\bm{x}_1, \bm{x}_2) &\sim \psi^\ast_i (t, \bm{x}_1) \chi^T_j (t, \bm{x}_2), \\
	\varphi^T_{ij}(t,\bm{x}_1, \bm{x}_2) &\sim \chi^\ast_i (t, \bm{x}_2) \psi^T_j (t, \bm{x}_1).
\end{align}	
\end{subequations}
When projecting the Hamiltonian we encounter following Pauli structures involving the bilocal field
\begin{subequations}
\begin{align}
	\varphi^\dagger_{ij}(t,\bm{x}_1, \bm{x}_2) \varphi_{ji}(t,\bm{x}_1, \bm{x}_2) &\sim 
	- \psi^\dagger_j (t, \bm{x}_1) \psi_j (t, \bm{x}_1) \chi^\dagger_i (t, \bm{x}_2) \chi_i (t, \bm{x}_2), \\
	\varphi^\dagger_{ij}(t,\bm{x}_1, \bm{x}_2) \bm{\sigma}_{jk} \varphi_{ki}(t,\bm{x}_1, \bm{x}_2) &\sim 
	- \psi^\dagger_j (t, \bm{x}_1) \bm{\sigma}_{jk} \psi_k (t, \bm{x}_1) \chi^\dagger_i (t, \bm{x}_2) \chi_i (t, \bm{x}_2), \\
	\varphi^\dagger_{ij}(t,\bm{x}_1, \bm{x}_2) \varphi_{jk}(t,\bm{x}_1, \bm{x}_2)  \bm{\sigma}_{ki} &\sim 
	-\psi^\dagger_j (t, \bm{x}_1) \psi_j (t, \bm{x}_1) \chi^\dagger_k (t, \bm{x}_2) \bm{\sigma}_{ki} \chi_i (t, \bm{x}_2).
\end{align}
\end{subequations}
Sometimes NREFT practitioners prefer to trade the field $\chi$ for $\chi_c$ defined as
\begin{equation}
	\chi_c = - i \sigma^2 \chi^\ast, \quad \chi_c^\dagger = i \chi^T \sigma^2,
\end{equation}
with
\begin{equation}
	\{ \chi_{c,i} (t, \bm{x}),  \chi_{c,j}^\dagger (t, \bm{x}') \} = \delta_{ij} \, \delta^{(3)} (\bm{x} - \bm{x}').
\end{equation}
The reason for this is that while the basis $(\psi,\chi)$ is best suited for studying annihilations and formations of heavy fermions, the set $(\psi,\chi_c)$ often turns out to be more convenient when dealing with scattering processes. Switching from $\chi$ to $\chi_c$ on the level of the pNREFT Lagrangian can be easily achieved via a suitable field redefinition, which amounts to introducing
\begin{subequations}
\begin{align}
	\tilde{\varphi}_{ij} (t,\bm{x}_1, \bm{x}_2) &=  -i \sigma^2_{ik} \varphi^T_{kl} (t,\bm{x}_1, \bm{x}_2) \sigma^2_{lj}   \sim 
	\chi_{c,i} (t,\bm{x}_2) \psi^T_j (t,\bm{x}_1),  \\
	\tilde{\varphi}^\dagger_{ij} (t,\bm{x}_1, \bm{x}_2) &= i  \varphi^\ast_{ik} (t,\bm{x}_1, \bm{x}_2) \sigma^2_{kj} \sim 	\psi^\ast_i (t,\bm{x}_1) \chi^\dagger_{c,j} (t,\bm{x}_2),
\end{align}
\end{subequations}
with
\begin{subequations}
\begin{align}
	\varphi^T_{ij} (t,\bm{x}_1, \bm{x}_2) & = i \sigma^2_{ik} \tilde{\varphi}_{kj} (t,\bm{x}_1, \bm{x}_2), \\
	\varphi^\ast_{ij} (t,\bm{x}_1, \bm{x}_2)  & =-i \tilde{\varphi}^\dagger_{ik} (t,\bm{x}_1, \bm{x}_2) \sigma^2_{kj}.
\end{align}
\label{eq:chitochic}
\end{subequations}
Then we have
\begin{subequations}
\begin{align}
	\varphi^\dagger_{ij}(t,\bm{x}_1, \bm{x}_2) \varphi_{ji}(t,\bm{x}_1, \bm{x}_2) &= 
	\tilde{\varphi}^\dagger_{ij}(t,\bm{x}_1, \bm{x}_2) \tilde{\varphi}_{ji}(t,\bm{x}_1, \bm{x}_2) \nonumber 	\\
	& \sim \psi^\dagger_j (t, \bm{x}_1) \psi_j (t, \bm{x}_1) \chi^\dagger_{c,i} (t, \bm{x}_2) \chi_{c,i} (t, \bm{x}_2) \\
	\varphi^\dagger_{ij}(t,\bm{x}_1, \bm{x}_2) \bm{\sigma}_{jk} \varphi_{ki}(t,\bm{x}_1, \bm{x}_2) &= \tilde{\varphi}^\dagger_{ik} (t,\bm{x}_1, \bm{x}_2) \bm{\sigma}_{ij} \tilde{\varphi}_{kj} (t,\bm{x}_1, \bm{x}_2) \nonumber \\
	& \sim \psi^\dagger_j (t, \bm{x}_1) \bm{\sigma}_{jk} \psi_k (t, \bm{x}_1) \chi^\dagger_{c,i} (t, \bm{x}_2) \chi_{c,i} (t, \bm{x}_2)  \\ 
	 \varphi^\dagger_{ij}(t,\bm{x}_1, \bm{x}_2) \varphi_{jk}(t,\bm{x}_1, \bm{x}_2)  \bm{\sigma}_{ki} & = 
	- \tilde{\varphi}^\dagger_{ij} (t,\bm{x}_1, \bm{x}_2) \bm{\sigma}_{jk} \tilde{\varphi}_{ki} (t,\bm{x}_1, \bm{x}_2) \nonumber \\
	& = \psi^\dagger_j (t, \bm{x}_1) \psi_j (t, \bm{x}_1) \chi^\dagger_{c,k} (t, \bm{x}_2) \bm{\sigma}_{ki} \chi_{c,i} (t, \bm{x}_2) ,
\end{align}
\end{subequations}
where in the first step we transposed the initial expression with respect to its Pauli indices and the employed \Eqs\eqref{eq:chitochic}. Furthermore, we used that
\begin{subequations}
\begin{align}
\chi_i^\dagger \chi_i &= - \chi_i^T \chi^\ast_i = - \chi_{c,i}^\dagger \chi_{c,i}, \\
\chi^\dagger_i \bm{\sigma}_{ij} \chi_j &= -\chi^T_j \bm{\sigma}_{ji}^T \chi^\ast_i = \chi_{c,k}^\dagger \bm{\sigma}_{kl} \chi_{c,l}
\end{align}
\label{eq:app-transp}
\end{subequations}
and
\begin{align}
\bm{\sigma}^T_{ji} & = - \sigma^2_{jk} \bm{\sigma}_{kl} \sigma^2_{li}.
\end{align}
In this context it is worth noting that, in general, one can eliminate $\chi$ in favor of  $\chi_c$ already at the level of the NREFT Lagrangian and then directly project onto
\begin{equation}
	\int d^3 \bm{x}_1   d^3 \bm{x}_2 \; \tilde{\varphi}_{ij}  (\bm{x}_1,\bm{x}_2) \psi^\dagger_i (\bm{x}_1) \chi^\dagger_{c,j}(\bm{x}_2) \ket{\phi_{\hbox{\tiny US}}}.
\end{equation}
In this case one would first apply \Eq\eqref{eq:app-transp} to the bilinear part of the NREFT Lagrangian and then make use of
the Fierz transformations in the Pauli space 
\begin{align}
	\delta_{ij} \delta_{lk} = \frac{1}{2} \delta_{il} \delta_{jk} + \frac{1}{2}  \bm{\sigma}_{li} \cdot \bm{\sigma}_{jk},  \\
	\bm{\sigma}_{ij} \cdot \bm{\sigma}_{kl} = \frac{3}{2} \delta_{il} \delta_{jk} - \frac{1}{2}  \bm{\sigma}_{li} \cdot \bm{\sigma}_{jk},
\end{align}
when rewriting the four-fermion operators accordingly.

Hence, even though physical results clearly do not depend on whether we use $\chi$ or $\chi_c$, the former turns out to
be more useful to study production or annihilation of heavy particles (as is often done in NREFTs), while the latter is very convenient for scattering processes typically arising in pNREFTs.

Finally, for the sake of completeness, let us provide the explicit $\mathcal{O}(1/M)$ result for the projection of $\mathcal{H}_{\gamma_5}$ onto the subspace spanned by \Eqs\eqref{eq:ket1} before the multipole expansion
\begin{align}
\braket{P|\mathcal{H}_{\gamma_5}|P'} &= \int d^3 \bm{x}_1  d^3 \bm{x}_2 \, \biggl [ g \varphi^\dagger_{ij}(t,\bm{x}_1, \bm{x}_2) \left ( \phi(t,\bm{x}_1) 
+ \phi(t,\bm{x}_2) \right ) \varphi_{ji}(t,\bm{x}_1, \bm{x}_2) \nonumber \\
& - \frac{( \bm{\nabla}_{\bm{x}_1}^2 \varphi^\dagger_{ij}(t,\bm{x}_1, \bm{x}_2))  \varphi_{ji}(t,\bm{x}_1, \bm{x}_2)}{2M} - \frac{\varphi^\dagger_{ij}(t,\bm{x}_1, \bm{x}_2) \bm{\nabla}_{\bm{x}_2}^2 \varphi_{ji}(t,\bm{x}_1, \bm{x}_2)}{2M} \nonumber \\
& + \frac{g_5^2}{2M} \varphi^\dagger_{ij}(t,\bm{x}_1, \bm{x}_2) \left ( \phi^2 (t,\bm{x}_1)
+ \phi^2 (t,\bm{x}_2) \right ) \varphi_{ji}(t,\bm{x}_1, \bm{x}_2) \nonumber \\
& - \frac{g_5}{2M} \varphi^\dagger_{ij}(t,\bm{x}_1, \bm{x}_2) 
\bm{\sigma}_{jk} \cdot (\bm{\nabla}_{\bm{x}_1}  \phi (t,\bm{x}_1))
\varphi_{ki}(t,\bm{x}_1, \bm{x}_2) \nonumber \\
& + \frac{g_5}{2M} \varphi^\dagger_{ij}(t,\bm{x}_1, \bm{x}_2) 
\bm{\sigma}_{ki} \cdot (\bm{\nabla}_{\bm{x}_2}  \phi (t,\bm{x}_2))
 \varphi_{jk}(t,\bm{x}_1, \bm{x}_2) \biggl ].
\label{eq:proj5}
\end{align}
Then we only need to switch to the center-of-mass coordinates via
\begin{align}
	\bm{x_{1/2}} = \bm{R} \pm \frac{1}{2} \bm{r}, \quad \bm{\nabla}_{\bm{x}_{1/2}} = \pm \bm{\nabla}_{\bm{r}} + \bm{\nabla}_{\bm{R}}
\end{align}
and multipole expand the scalar fields in $\bm{r}$ up to the desired order using
\begin{align}
	\phi (t, \bm{x}_{1/2} ) &= \phi\left(t, \bm{R} \pm \frac{\bm{r}}{2} \right) = \phi(t,\bm{R}) \pm \frac{\bm{r}}{2} \cdot \bm{\nabla} \phi (t,\bm{R}) + \frac{1}{8} r^i r^j \nabla_R^i \nabla_R^j \, \phi (t,\bm{R}) + \mathcal{O}(r^3)
\end{align}
to arrive at the final bilinear piece of the pNRY$_{\gamma_5}$ Lagrangian at $\mathcal{O}(1/M)$.

\section{Matrix elements}
\label{sec:appendix-me}
The matrix elements presented in the main body of the paper can be evaluated analytically in the Coulomb limit,
where we essentially employ analytic solutions for the bound and scattering state wave functions of the Hydrogen atom.
The bound state wave function reads
\begin{equation}
	\psi_{n \ell m}(\bm{r}) = R_{n\ell}(r) Y_{\ell m} (\theta,\phi) \, , 
\end{equation}
with
\begin{align}
	R_{n\ell}(r) &= \frac{1}{(2 \ell +1)!}\sqrt{\left( \frac{2}{n a_0}\right)^3  \frac{(n+\ell)!}{2n(n-\ell-1)!}} \left( \frac{2 r }{n a_0}\right)^{\ell} \nonumber \\
	& \times {}_1 F_1 \left( \ell +1 -n; 2 \ell +2; \frac{2 r }{n a_0} \right)\, e^{-r/(n a_0)},
\end{align}
where $Y_{\ell m} (\theta,\phi)$ is the spherical harmonic, while ${}_1 F_1(a;b;z)$ denotes the confluent hypergeometric or Kummer's function. The scattering wave functions arise as solutions of the Schrödinger equation
\begin{equation}
	\left (-  \frac{\nabla^2}{2m} + \frac{Z \alpha}{r} \right ) \psi_{\bm{p}} (\bm{r}) = \frac{p^2}{2m} \psi_{\bm{p}} (\bm{r}),
\end{equation}
with $Z = -1$ for the Hydrogen atom. Solving this equation in the parabolic coordinates and choosing the solution that describes the wave before approaching the source at the origin we can expand the resulting wave function in plane waves. This yields
\begin{align}
	\psi_{\bm{p}} (\bm{r}) &= \sqrt{\frac{2 \pi \zeta}{1-e^{-2 \pi \zeta}}}  e^{ipr}  \sum_{\ell=0}^{\infty} {e^{i \sigma_\ell}} e^{{+}i \frac{\pi \ell}{2}} \frac{(2 p r)^\ell}{(2 \ell)!} \phantom{x}_1 F_1 \left( \ell +1 -i \zeta; 2 \ell +2; -2 i pr  \right) \nonumber \\
	& \times P_\ell (\cos \theta)\prod_{s=1}^{\ell} \left( s^2+\zeta^2 \right)^{\frac{1}{2}},
	\label{scattering_wave_function}
\end{align}
where $\zeta \equiv \alpha / v = - \eta$ is related to the Sommerfeld parameter $\eta$, while $\sigma_l$ denotes the Coulomb phase shift
\begin{equation}
	\sigma_l = \arg \Gamma(l+1 + i \eta),
\end{equation}
that satisfies
\begin{equation}
	e^{2 i \sigma_l} = \frac{\Gamma(1+l + i \eta)}{\Gamma(1+l - i \eta)}.
\end{equation}
Notice that the scattering state wave function is decomposed in terms of partial waves with definite angular momentum $\ell$ and that we choose the relative momentum $\bm{p}$ to be aligned along the $z$-axis. This agrees with the conventions commonly used in quantum mechanics textbooks~\cite{merzbacher1998quantum,messiah1999quantum} and DM literature~\cite{Wise:2014jva,Oncala:2018bvl}.

\subsection{Expectation values of \texorpdfstring{$\bm{r}^2$}{r 2}}

We start with the discussion of matrix elements involving the position operator squared
\begin{equation}
  \langle\bm{p} | \bm{r}^2 |n \ell m \rangle = \sum_{\ell'} \int d^3 \bm{r} \, r^2 \psi^*_{\ell'}(\bm{r})  \psi_{n \ell m} (\bm{r}).
\end{equation}
This integral obviously factorizes into a radial and an angular part. The integration over the solid angle
can be straightforwardly evaluated using the following identity
\begin{equation}
	\int d \Omega \; P^*_{\ell'} (\cos \theta) Y_{\ell m} (\theta,\phi)  = \left(\frac{4 \pi}{2 \ell' +1} \right)^{1/2}  \int d \Omega \; Y_{\ell',0} Y_{\ell m}   = \left(\frac{4 \pi}{2 \ell' +1} \right)^{1/2}  \delta_{\ell \ell'} \delta_{m,0} \, .
	\label{eq:angular-part}
\end{equation}
The remaining radial integral\footnote{
	Notice that ${_1}F_1^\ast (\ell +1 -i \zeta; 2 \ell +2; -2 i pr )= {_1}F_1(\ell +1 +i \zeta; 2 \ell +2; 2 i pr )$ for  $p,\zeta \in \mathbb{R}$.}
\begin{equation}
	\int_0^\infty dr \; r^{4 + 2 \ell} \, e^{-\frac{r}{2} \left( 2 i p + \frac{2}{n a_0}\right)}     \phantom{\,}_1 F_1 \left( \ell +1 +i \zeta; 2 \ell +2; +2 i pr  \right) \phantom{x}_1 F_1 \left( \ell +1 -n; 2 \ell +2; \frac{2 r }{n a_0} \right)
	\label{radial_integral_scalar}
\end{equation}
for generic quantum numbers $n$ and $\ell$ can be calculated with the aid of the relation (\cf  section~7.622 in~\cite{TSintegrals}) 
\begin{align}
	&\int_0^\infty dt \, t^{c-1} e^{- \rho t} \phantom{\,}_1 F_1 \left(a , c ;t  \right) \phantom{\,}_1 F_1 \left(b , c ; \lambda t  \right) \nonumber  \\
	& = \Gamma(c) (\rho-1)^{-a} (\rho-\lambda)^{-b}  \rho^{a+b-c} {}_2 F_1 \left(a,b , c ; \lambda (\rho-1)^{-1} (\rho-\lambda)^{-1}\right) \, ,
	\label{hyper_aid}
\end{align}
where we set
\begin{align}
	a &=\ell+1-n \, , \quad b = \ell +1 + i \zeta \, , \quad c = 2\ell +2 \, , \nonumber \\
	\rho&=\frac{1}{2} \left( 1+ i\frac{n}{\zeta} \right) \, , \quad \lambda= i\frac{n}{\zeta} \, ,
	\quad t=\frac{2r}{n a_0}.
	\label{eq:hyper_values}
\end{align}
Differentiating both sides of \Eq\eqref{hyper_aid} with respect to $\rho$ three times and plugging
in the values from \Eq\eqref{eq:hyper_values} we can directly evaluate the radial integral and arrive 
at a compact analytic expression for the desired expectation value
\begin{align}
	\sum_{\ell'}   \langle\bm{p} \ell'| \bm{r}^2 |n \ell 0\rangle &={e^{-i\sigma_\ell}}(-1)^{n-\ell} \delta_{m,0}  \, 2 \pi \, \left( \frac{n a_0}{2}\right)^{\frac{7}{2}} \frac{2^{8 +{2}\ell}\Gamma(2\ell+2)}{(2 \ell +1)! (2 \ell)!}\sqrt{ \frac{n (n+\ell)!}{(2\ell+1)(n-\ell-1)!}} \nonumber \\
	& \times \prod_{s=1}^{\ell} \left( s^2+\zeta^2 \right)^{\frac{1}{2}}
	n^{\ell } \zeta^{6+\ell} \sqrt{\frac{ \zeta}{1-e^{-2 \pi \zeta}}}  \frac{e^{-2 \zeta\arccot(\zeta/n)}}{(n^2+\zeta^2)^{3+\ell}} e^{-i2(n-\ell -1)  \arccot(\zeta/n)} 
	\nonumber\\ 
	& \times e^{-i \frac{\pi \ell}{2}} {}_2 F_1 \left(1+\ell -n, 1+\ell +i \zeta, 2\ell +2  ;\frac{4in\zeta}{(n+i \zeta)^2} \right).
	\label{eq:r2master}
\end{align}
Notice that the arising phase cancels out when squaring the above quantity.\footnote{It should be noted, that for $nS$ states, the remaining phase $e^{-2i(n-1)\arccot(\zeta/n)}$ is exactly canceled by the factor that appears in $_2 F_1(1-n,1+i \zeta,2,4in\zeta/(n+i\zeta)^2)$ when setting $n$ to specific integer values. The matrix element is then purely real.} Let us also remark that if  we are interested only in the expectation values for fixed quantum numbers (\eg $n=1, \ell=0, m=0$ etc.), then the radial integral in \Eq\eqref{radial_integral_scalar} simplifies and can be evaluated using a much simpler identity (\cf section~7.522 in~\cite{TSintegrals}) that reads
\begin{equation}
	\int_0^\infty dt \, t^{b-1} \, e^{-zt} \phantom{a}_1 F_1 (a,c;kt) = \Gamma(b) z^{-b} \,{}_2 F_1(a,b,c;k/z) \, ,
	\label{eq:hyper1}
\end{equation}
with $\textrm{Re}(b)>0 $ and $\textrm{Re}(z)> \textrm{max}(\textrm{Re}(k),0) $.

Finally, let us provide explicit results for the phenomenologically relevant lowest-lying $S$- and $P$-wave matrix elements. The matrix elements squared for the first three $nS$ states read
\begin{subequations}
	\begin{align}
		| \bra{\bm{p}} \bm{r}^2 \ket{1S}|^2 &= \frac{2^{18} \pi^2}{\, M^7 \alpha^7} \left( \frac{\zeta}{1-e^{-2 \pi \zeta}}  \right)   \frac{\zeta^{12}}{ (1+\zeta^2)^6 } e^{-4 \zeta \arccot(\zeta)}  \, ,
		\\
		| \bra{\bm{p}} \bm{r}^2 \ket{2S}|^2 &= \frac{2^{27} \pi^2}{ \, M^7 \alpha^7} \left( \frac{\zeta}{1-e^{-2 \pi \zeta}}  \right)    \frac{\zeta^{12} (4+3 \zeta^2)^2}{(4+\zeta^2)^8} e^{-4 \zeta  \arccot(\zeta/2)} \, ,
		\\
		| \bra{\bm{p}} \bm{r}^2 \ket{3S}|^2 &= \frac{2^{18} 3^9 \pi^2}{ \, M^7 \alpha^7} \left( \frac{\zeta}{1-e^{-2 \pi \zeta}}  \right)    \frac{\zeta^{12} (81+78\zeta^2 +13\zeta^4)^2}{(9+\zeta^2)^{10}} e^{-4 \zeta  \arccot(\zeta/3)},
	\end{align}
\end{subequations}
where $\ket{1S}$, $\ket{2S}$ and $\ket{3S}$ should be understood as $\ket{100}$, $\ket{200}$ and $\ket{300}$ respectively. In the case of $P$-waves we consider $\ket{210}$, $\ket{21 \pm1}$, $\ket{310}$ and $\ket{31 \pm1}$ where only  the states with $m=0$ are non-vanishing. We obtain
\begin{subequations}
	\begin{align}
		| \bra{\bm{p}} \bm{r}^2 \ket{210}|^2 &= \frac{2^{31} \pi^2}{\, M^7 \alpha^7} \left( \frac{\zeta}{1-e^{-2 \pi \zeta}}  \right)   \frac{\zeta^{14}(1+\zeta^2)}{ (4+\zeta^2)^8 } e^{-4 \zeta \arccot(\zeta/2)}  \, ,
		\\
		| \bra{\bm{p}} \bm{r}^2 \ket{310}|^2 &= \frac{2^{23} 3^{10} \pi^2}{\, M^7 \alpha^7} \left( \frac{\zeta}{1-e^{-2 \pi \zeta}}  \right)   \frac{\zeta^{14}(1+\zeta^2)(9+2 \zeta^2)^2}{ (9+\zeta^2)^{10} } e^{-4 \zeta \arccot(\zeta/3)} \, .
	\end{align}
\end{subequations}
We have explicitly verified these results by calculating the matrix elements either via the master formula 
\Eq\eqref{eq:r2master} or on a case-by-case basis using \Eq	\eqref{eq:hyper1}. Furthermore, 
we conducted a numerical check that gave a perfect agreement over a large range of $\zeta$ values.

\subsection{Expectation values of \texorpdfstring{$\bm{\nabla}^2$}{del 2}}

As far as the evaluation of the matrix element
\begin{equation}
  \langle\bm{p}| \nabla^2_{\bm{r}}/M^2 |n \ell m \rangle = \frac{1}{M^2} \sum_{\ell'} \int d^3 \bm{r} \, \psi^*_{\ell'}(\bm{r}) \nabla^2_{\bm{r}}  \psi_{n \ell m} (\bm{r})
\end{equation}
is concerned, we choose to proceed on the case-by-case basis by considering explicit values of the quantum numbers $n$, $\ell$ and $m$. To this end we only need to use the identities given in 
\Eq\eqref{eq:angular-part} and \Eq\eqref{eq:hyper1}. This yields
\begin{subequations}
	\begin{align}
		\langle\bm{p}| \nabla^2_{\bm{r}}/M^2 |1S\rangle &= -2^3 {e^{-i\sigma_0}} \pi \sqrt{\frac{\zeta}{1-e^{- 2 \pi \zeta}}} \frac{\sqrt{M \alpha}}{M^2 } \frac{\zeta^2}{(1+\zeta^2)} e^{-2\zeta \arccot(\zeta )} \, ,
		\\
		\langle\bm{p}| \nabla^2_{\bm{r}}/M^2 |2S\rangle &= {-}2^{7/2} {e^{-i\sigma_0}}\pi \sqrt{\frac{\zeta}{1-e^{- 2 \pi \zeta}}} \frac{\sqrt{M \alpha}}{M^2 } \frac{\zeta^2 (4+3 \zeta^2)}{(4+\zeta^2)^2} e^{-2\zeta \arccot(\zeta /2)} \, ,
		\\
		\langle\bm{p}| \nabla^2_{\bm{r}}/M^2 |3S\rangle &= -2^{3} 3^{1/2} {e^{-i\sigma_0}} \pi \sqrt{\frac{\zeta}{1-e^{- 2 \pi \zeta}}} \frac{\sqrt{M \alpha}}{M^2 } \frac{\zeta^2 (81+78 \zeta^2+13 \zeta^4)}{(9+\zeta^2)^3} e^{-2\zeta \arccot(\zeta /3)} \, , \\
		\langle\bm{p}| \nabla^2_{\bm{r}}/M^2 |210\rangle &= i \,  2^{11/2} \pi \sqrt{\frac{\zeta}{1-e^{- 2 \pi \zeta}}} \frac{\sqrt{M \alpha}}{M^2 } \frac{\zeta^3 (1+i \zeta) }{(4+\zeta^2)^2} e^{-2\zeta \arccot(\zeta /2)} \, ,
		\\
		\langle\bm{p}| \nabla^2_{\bm{r}}/M^2 |310\rangle &= i \, 3 \cdot 2^{11/2}   \pi \sqrt{\frac{\zeta}{1-e^{- 2 \pi \zeta}}} \frac{\sqrt{M \alpha}}{M^2 } \frac{\zeta^3 (1+i \zeta) (2 \zeta^2+9) }{(9+\zeta^2)^3} e^{-2\zeta \arccot(\zeta /3)},
	\end{align}
\end{subequations}
where we would like to remark that matrix elements for the $P$-wave states $\ket{21\pm1}$ and $\ket{31\pm1}$ vanish. The so-obtained results are also in perfect agreement with our numerical checks.

\subsection{Expectation values of \texorpdfstring{$r^i r^j $}{ri rj}}

Last but not least, we also need to deal with the tensor matrix elements given by
\begin{equation}
	  \langle\bm{p} | r^i r^j |n \ell m \rangle = \frac{1}{M^2} \sum_{\ell'} \int d^3 \bm{r} \, r^i r^j \psi^*_{\ell'}(\bm{r})  \psi_{n \ell m} (\bm{r}).
\end{equation}
Here it is useful to observe that the first 9 spherical harmonics $Y_{lm}$ with $\ell\leq2$ and the corresponding $m$-values can be written as linear combinations of $x$, $y$, $z$, $x^2$, $y^2$, $z^2$, $x y$, $x z$ and $y z$. Inverting this linear system we can express each of these 9 quantities as a linear combination of the spherical harmonics. More explicitly, introducing $T^{ij} = r^i r^j$ we have
\begin{subequations}
	\begin{align}
		T^{xx} &= r^2 \left[ \frac{2\sqrt{ \pi}}{3} Y_{0,0} - \frac{2}{3} \sqrt{\frac{ \pi}{5}} Y_{2,0} + \sqrt{\frac{2 \pi}{15}} (Y_{2,2}+Y_{2,-2}) \right] \, , 
		\label{eq:tensor-spherical_a}
		\\
		T^{yy} &= r^2 \left[\frac{2\sqrt{ \pi}}{3} Y_{0,0} - \frac{2}{3} \sqrt{\frac{ \pi}{5}} Y_{2,0} - \sqrt{\frac{2 \pi}{15}} (Y_{2,2}+Y_{2,-2}) \right] \, , 
		\\
		T^{zz} &= r^2\left( \frac{2\sqrt{ \pi}}{3} \, Y_{0,0}+  \frac{4}{3} \sqrt{ \frac{ \pi}{5}} \, Y_{2,0} \right) \, ,   \quad  T^{xy} = i r^2 \sqrt{\frac{2 \pi}{15}} \left( Y_{2,-2}-Y_{2,2} \right)
		\\
		T^{xz} &=  r^2 \sqrt{\frac{2 \pi}{15}} \left( Y_{2,-1}-Y_{2,1} \right) \, , \quad  T^{yz} = i r^2 \sqrt{\frac{2 \pi}{15}} \left( Y_{2,1}+Y_{2,-1} \right)\,.
		\label{eq:tensor-spherical_d}
	\end{align}
\end{subequations}
Notice that while the above results can be found in many quantum mechanics textbooks (\cf \eg\cite{merzbacher1998quantum}), the algorithm itself naturally works also for higher rank tensors beyond
rank 2. Applying \Eqs\eqref{eq:tensor-spherical_a}--\eqref{eq:tensor-spherical_d} to our problem leaves us with angular integrals over products of three spherical harmonics. Those can be evaluated using
\begin{align}
	& \int d \Omega \; [Y_{\ell_3,m_3}(\theta,\phi)]^* \, Y_{l_1,m_1}(\theta,\phi) \,  Y_{l_2,m_2}(\theta,\phi) \nonumber \\
	&= \sqrt{\frac{(2 \ell_1+1)(2 \ell_2+1)}{(4 \pi) (2 \ell_3+1)}} \langle \ell_1 \ell_2 00| \ell_1 \ell_2 \ell_3 0 \rangle \langle \ell_1 \ell_2 m_1 m_2| \ell_1 \ell_2 \ell_3 m_3 \rangle \,,
	\label{integral_3_sp_harm}
\end{align}
where the quantities in the angle brackets are the familiar angular momentum Clebsch-Gordan coefficients. As far as the radial integrals are concerned, \Eq\eqref{eq:hyper1} again turns out to be sufficient for all cases with explicit values of the quantum number $n$, $\ell$ and $m$.

Putting everything together, we obtain following expressions for the first three $S$-wave states
\begin{subequations}
	\begin{align}
		| \bra{\bm{p}} r^i r^j \ket{1S}|^2 &=\frac{2^{18} \pi^2}{3 \, M^7 \alpha^7} \left( \frac{\zeta}{1-e^{-2 \pi \zeta}}  \right) \frac{\zeta^{12} ( 33+ 9 \zeta^2  )}{ (1+\zeta^2)^7 }  e^{-4\zeta \arccot(\zeta)}  \, , \\
		|\bra{\bm{p}} r^i r^j \ket{2S}|^2 &=\frac{2^{27} \pi^2}{3 \, M^7 \alpha^7} \left( \frac{\zeta}{1-e^{-2 \pi \zeta}}  \right) \frac{\zeta^{12} \left[ (4+\zeta^2)(4+3\zeta^2)^2+2048(1+\zeta^2) \right] }{ (4+\zeta^2)^9 } \nonumber \\
		& \times e^{-4 \zeta \arccot(\zeta/2)}  \, ,	\\
		|\bra{\bm{p}} r^i r^j \ket{3S}|^2 &=\frac{2^{18} 3^8 \pi^2}{ \, M^7 \alpha^7} \left( \frac{\zeta}{1-e^{-2 \pi \zeta}}  \right) \frac{e^{-4 \zeta \arccot(\zeta/3)}}{ (9+\zeta^2)^{10} } \zeta^{12} \nonumber \\
		& \times \left[ 2^{3} 3^{2} (\zeta^2-27)^2(\zeta^4+5\zeta^2+4) +(13 \zeta^4 +78 \zeta^2 +81)^2  \right] \, .
	\end{align}
\end{subequations}
In the case of $P$-wave states we find
\begin{subequations}
	\begin{align}
		|\bra{\bm{p}} r^i r^j \ket{210}|^2 &=\frac{2^{31} \pi^2}{ \, M^7 \alpha^7} \left( \frac{\zeta}{1-e^{-2 \pi \zeta}}  \right) \frac{\zeta^{14} (\zeta^2 +1)(3 \zeta^2+44) }{ (4+\zeta^2)^9 } e^{-4 \zeta \arccot(\zeta/2)}  \, , \
		\\
		|\bra{\bm{p}} r^i r^j \ket{310}|^2 &=\frac{2^{23} \, 3^{10} \pi^2}{ \, M^7 \alpha^7} \left( \frac{\zeta}{1-e^{-2 \pi \zeta}}  \right) \frac{\zeta^{14} (\zeta^2 +1)(2 \zeta^6 + 108 \zeta^4 + 2349 \zeta^2 + 8019 ) }{ (9+\zeta^2)^{11} } \nonumber \\
		& \times e^{-4 \zeta \arccot(\zeta/3)}  \, .
	\end{align}
\end{subequations}
At variance with the results for the other two types of matrix elements involving $r^2$ and $\bm{\nabla}^2$, here we also find nonvanishing matrix elements squared for $m \neq 0$
\begin{subequations}
	\begin{align}
		|\bra{\bm{p}} r^i r^j \ket{21 \pm 1}|^2 &=\frac{2^{32} \pi^2}{ \, M^7 \alpha^7} \left( \frac{\zeta}{1-e^{-2 \pi \zeta}}  \right) \frac{\zeta^{14} (\zeta^2 +1) }{ (4+\zeta^2)^8 } e^{-4 \zeta \arccot(\zeta/2)}  \, , \
		\\
		|\bra{\bm{p}} r^i r^j \ket{31 \pm 1}|^2 &=\frac{2^{24} \, 3^{10} \pi^2}{ \, M^7 \alpha^7} \left( \frac{\zeta}{1-e^{-2 \pi \zeta}}  \right) \frac{\zeta^{14} (\zeta^2 +1)( \zeta^4 +27 \zeta^2 + 81 ) }{ (9+\zeta^2)^{10} } e^{-4 \zeta \arccot(\zeta/3)}.
	\end{align}
\end{subequations}

 \section{Bound-state dissociation cross sections} 
\label{app:diss_excited}
In section~\ref{sec:bsd} we provided only one explicit result ($\ket{1S}$ state) for the bound-state dissociation cross section. For the sake of completeness, here we list the results for the remaining $S$-
and $P$-wave states that can be calculated using the matrix elements given in 
appendix~\ref{sec:appendix-me}. The bound-state dissociation cross sections for the $\ket{2S}$ and $\ket{3S}$ states read
\begin{subequations}
\begin{align}
	\sigma^{2S}_{\textrm{bsd}}(|\bm{k}|)  &=  \frac{ \alpha^3 \, 2^{8} \pi^2}{15} \frac{|E_2|^2}{M|\bm{k}|^3} \left( 7 + 12 \frac{|E_2|}{|\bm{k}|} -20 \frac{|E_2|^2}{|\bm{k}|^2} + 16 \frac{|E_2|^3}{|\bm{k}|^3}\right)\frac{e^{-\frac{4}{w_2(|\bm{k}|)} \arctan (2 w_2(|\bm{k}|))}}{1-e^{-\frac{2 \pi}{ w_2(|\bm{k}|) }}} \, ,	\\
	\sigma^{3S}_{\textrm{bsd}}(|\bm{k}|)  &=   \frac{ \alpha^3 \, 2^{7} \pi^2}{45}  \frac{|E_3|^2}{M|\bm{k}|^3} \left( 63 + 516 \frac{|E_3|}{|\bm{k}|} +208 \frac{|E_3|^2}{|\bm{k}|^2} 
	 + 576 \frac{|E_3|^3}{|\bm{k}|^3} +1280 \frac{|E_3|^4}{|\bm{k}|^4}\right) \nonumber \\
	& \times \frac{e^{-\frac{4}{w_3(|\bm{k}|)} 
	 \arctan (3 w_3(|\bm{k}|))}}{1-e^{-\frac{2 \pi}{ w_3(|\bm{k}|) }}} \, ,
\end{align}
\end{subequations}
where $w_n(|\bm{k}|) \equiv \sqrt{|\bm{k}|/|E_n|-1} /n$, $E_n = - M \alpha^2 / (4 n^2)$ and
$\zeta = 1/w_n(|\bm{k}|)$. For the $P$-wave cross sections we find
\begin{subequations}
	\begin{align}
		\sigma^{2P}_{\textrm{bsd}}(|\bm{k}|)  &=  \frac{ \alpha^3 \, 2^{10} \pi^2}{15} \frac{|E_2|^3}{M|\bm{k}|^4} \left( 9 + 23 \frac{|E_2|}{|\bm{k}|} - 12 \frac{|E_2|^2}{|\bm{k}|^2} \right)\frac{e^{-\frac{4}{w_2(|\bm{k}|)} \arctan (2 w_2(|\bm{k}|))}}{1-e^{-\frac{2 \pi}{ w_2(|\bm{k}|) }}} \, ,	\\
		\sigma^{3P}_{\textrm{bsd}}(|\bm{k}|)  &=   \frac{ \alpha^3 \, 2^{12} \pi^2}{15}  \frac{|E_3|^2}{M|\bm{k}|^4} \left(9 + 75 \frac{|E_3|}{|\bm{k}|} + 17 \frac{|E_3|^2}{|\bm{k}|^2} 
		- 52\frac{|E_3|^3}{|\bm{k}|^3} + 32 \frac{|E_3|^4}{|\bm{k}|^4}\right) \nonumber \\
		& \times \frac{e^{-\frac{4}{w_3(|\bm{k}|)} 
				\arctan (3 w_3(|\bm{k}|))}}{1-e^{-\frac{2 \pi}{ w_3(|\bm{k}|) }}} \, .
	\end{align}
\end{subequations}

\bibliographystyle{hieeetr}
\bibliography{paper.bib}

\end{document}